\documentclass[journal]{IEEEtran}

\usepackage{lipsum}
\usepackage{graphicx}
\ifCLASSOPTIONcompsoc
    \usepackage[caption=false, font=normalsize, labelfont=sf, textfont=sf]{subfig}
\else
\usepackage[caption=false, font=footnotesize]{subfig}
\fi

\usepackage{graphicx}
\usepackage{amsfonts}
\usepackage{amsmath}
\usepackage{bm}
\usepackage{cite}
\usepackage{epstopdf}
\usepackage{booktabs}
\usepackage{lipsum}
\usepackage{balance}
\usepackage{amssymb}
\usepackage{ntheorem}
\usepackage[ruled,linesnumbered]{algorithm2e}
\usepackage{multirow}
\usepackage{color}
\theorembodyfont{\rmfamily}

\newtheorem{remark}{Remark}

\begin{document}

\title{Transmit Beamspace DDMA Based Automotive MIMO Radar} 

\author{Feng Xu, \IEEEmembership{Student Member, IEEE}, Sergiy A. Vorobyov, \IEEEmembership{Fellow, IEEE}, and Fawei Yang, \IEEEmembership{Student Member, IEEE}
\thanks{This work was supported in part by Huawei, and in part by the China Scholarship Council. This work was conducted while Feng Xu was a visiting doctoral student with the Department of Signal Processing and Acoustics, Aalto University. (\textit{Corresponding author: Feng Xu.})}
\thanks{Feng Xu is with the School of Information and Electronics, Beijing Institute of Technology, Beijing 100081, China, and also with the Department of Signal Processing and Acoustics, Aalto University, Espoo 02150, Finland. (e-mail: fengxu@bit.edu.cn, feng.xu@aalto.fi).}
\thanks{Sergiy A. Vorobyov is with the Department of Signal Processing and Acoustics, Aalto University, Espoo 02150, Finland. (e-mail: svor@ieee.org).}
\thanks{Fawei Yang is with the School of Information and Electronics, Beijing Institute of Technology, Beijing 100081, China. (e-mail: yangfawei@bit.edu.cn).}
}
\maketitle

\begin{abstract}
The time division multiple access (TDMA) technique has been applied in automotive multiple-input multiple-output (MIMO) radar. However, it suffers from the transmit energy loss, and as a result the parameter estimation performance degradation when the number of transmit elements increases. To tackle these problem, a transmit beamspace (TB) Doppler division multiple access (DDMA) approach is proposed. First, a phase modulation matrix with empty Doppler spectrum is introduced. By exploiting the empty Doppler spectrum, a test function based on sequential detection is developed to mitigate the Doppler ambiguity in DDMA waveform. Then, a discrete Fourier transform (DFT)-based TB in slow-time is formed. 
The proposed method can achieve waveform diversity in Doppler domain and generate a TB in slow-time that concentrates the transmitted power in a fixed spatial region to improve the transmit energy distribution for automotive MIMO radar, which is favored by medium/long range radar (MRR/LRR) applications. As compared to the conventional TDMA technique, the proposed TB DDMA approach can fully exploit the transmission capabilities of all transmit elements to ensure that the emitted power is efficiently used and inherits easy implementation. Moreover, the proposed TB DDMA method avoids the 
trade-off between the active time for each transmit antenna and the frame time. Simulation results verify the effectiveness of the proposed TB DDMA approach for automotive MIMO radar.
\end{abstract}


\section{Introduction}
\IEEEPARstart{R}{ecently}, automotive radar has received a lot of attention in the area of advanced driver assistance system (ADAS) due to its all weather, day and night surveillance ability \cite{1,2,29}. Together with other sensors like lidar, optical camera and ultrasound, automotive radar collects a wealth of information on the surrounding environment in multiple dimensions including range, Doppler and angle, which builds the cornerstone of autonomous driving. The automobile industry is quite familiar with radar technique, e.g., the automobile manufacturers have been using radar on parking assistance system for decades \cite{3}.

Frequency modulated continuous wave (FMCW) radar has dominated the automotive radar market over the last decades \cite{4,5}, and it has been updated to the multiple-input multiple-output (MIMO) radar configuration by exploiting time division multiple access (TDMA) technique \cite{6,7,8}. For example, Texas Instruments has developed a mmWave MIMO radar sensor based on low-power 45~nm RFCMOS technology \cite{8}, which typically has three transmit elements and four receive elements. Based on the MIMO radar concept, the spatial resolution can be  improved by increasing the number of antenna elements in the virtual array (VA), which is ensured by the orthogonality among transmitted waveforms \cite{9}. Although there exists several waveform design methods like frequency division multiple access (FDMA) and code division multiple access (CDMA) \cite{6,10}, the TDMA technique is mainly used due to its low cost and easy implementation. Moreover, TDMA MIMO radar can be easily upgraded from its single-input multiple-output (SIMO) counterparts, which is favored by the manufacturers.

The significant disadvantage of the TDMA technique is, however, the parameter estimation performance loss \cite{6}. Indeed, since only one antenna element is active at any given time, the transmitted energy is limited, which shortens the target detection range. If the transmit beampattern is omnidirectional, the power that can be reflected by potential targets is also confined as most of the energy is wasted. The target detection performance then degrades. Moreover, the TDMA technique requires sufficient pulse repetition frequency (PRF) tolerance, which is difficult to achieve in automotive radar due to the tight constrains between the maximum detectable range and the maximum unambiguous velocity.

To tackle the above listed problems, the transmit beamspace (TB) technique  \cite{11,12,13} can be used. By exploiting the TB matrix, the transmitted energy of automotive MIMO radar can be concentrated within a fixed spatial region, that enables to compensate the power loss for potential targets. The parameter estimation performance can be therefore improved. Furthermore, the use of the TB technique is quite suitable for automotive radar application since the demanded azimuthal field of view (FoV) in medium/long range radar (MRR/LRR) is physically confined \cite{1}. The energy emitted to the spatial region with large azimuth is wasted.

It is worth noting that some alternatives to FMCW radar have been proposed and investigated in \cite{14,15,16}. Code families like Gold, Barker and M-sequence have been utilized to achieve waveform diversity while keeping the full transmit capacity. Moreover, a coding method named Doppler division multiple access (DDMA) has been implemented in automotive radar \cite{17,18}. Historically, the DDMA approach was first used in airborne platforms for ground moving target indication (GMTI) \cite{19,20,21,22}. Instead of achieving waveform diversity in time domain, DDMA enables the design of orthogonal waveform in Doppler domain. More importantly, DDMA MIMO radar inherits the property of TDMA regarding the transmitter structure. Then the upgrade required is only the addition of a sequence of phase shifters, which makes the implementation acceptably simple and practical. The problem of transmitted energy loss can also be mitigated since the transmit elements can emit simultaneously. Overall, considering the performance to cost ratio, the DDMA technique is very promising for automotive MIMO radar.

Despite the above mentioned advantages, the DDMA approach has some drawbacks which limit its application in automotive radar. Since the waveform diversity is achieved in Doppler domain by modulating each transmit element with a distinct Doppler shift, the unambiguous Doppler region after matched-filtering is reduced. The maximum detectable velocity is also declined accordingly. To avoid this, the time duration of the chirp signal has to be decreased. As a consequence, the DDMA approach is only suitable for short range radar (SRR) applications. Meanwhile, similar to the TDMA approach, the DDMA approach also suffers from the waste of finite transmit energy as it generates an omnidirectional beampattern in slow-time. To fully exploit the capabilities of the DDMA approach for automotive MIMO radar application, it is necessary to overcome these disadvantages.

In this paper, a TB DDMA approach is proposed in application to automotive MIMO radar for improving transmit energy distribution. Instead of using the entire Doppler spectrum, the phase modulation matrix in the TB DDMA approach is designed to preserve a part of the Doppler spectrum. The empty Doppler spectrum enables a sequential detection that matches the default Doppler shifts, which the received signal has experienced, with the transmit elements in the correct order. The maximum unambiguous target Doppler shift can therefore be recovered. By exploiting the virtual transmit beamforming property of the DDMA waveform, a discrete Fourier transform (DFT)-based TB can be formed in slow-time via pulse selection in order to focus the transmitted energy within a specific region of interest. The waveform diversity in Doppler domain is maintained. As compared to the TDMA technique, the proposed TB DDMA approach fully exploits the transmission capabilities of the transmit elements and avoids the trade-off between the active time for each transmit antenna and the frame time. The proposed method can also be upgraded from a SIMO radar configuration easily by just adding a sequence of phase shifters. Simulation results verify the validity of the proposed TB DDMA approach for automotive MIMO radar.

{\it Notation}: Scalars, vectors, and matrices are represented by lower-case, boldface lower-case and boldface upper-case letters, e.g., $a$, $\bf a$ and $\bf A$, respectively. The transposition, Hermitian transposition and conjugate operators are denoted by ${\left( \cdot  \right)^T},{\left(  \cdot  \right)^H}$ and $*$, respectively, while $| \cdot |$ stands for a magnitude of a complex number or an absolute value of a real number. The notation $diag({\bf{a}})$ returns a diagonal matrix built out of a vector argument and $\bigcap$ denotes the logic AND operator, and $\| \cdot \|$ denotes the Euclidian norm of a vector. Moreover, ${{\bf{I}}_{M}}$ denotes the identity matrix of dimension $M \times M$, while ${\bf J}_M$ represents a square matrix with ones on its anti-diagonal and zeros elsewhere, i.e., the exchange matrix. For ${\bf A} \in {{\mathbb{C}}^{M \times N}}$, the $n$-th column vector, $(m,n)$-th element and the trace are denoted by ${\bf a}_n$, $a_{m,n}$ and $tr\{ {\bf A}\}$, respectively.

\section{Signal Model and Preliminaries}
\begin{figure}
    \centering
  \subfloat[Simplified diagram for FMCW signal generator.\label{fig1a}]{%
       \includegraphics[width=0.8\linewidth]{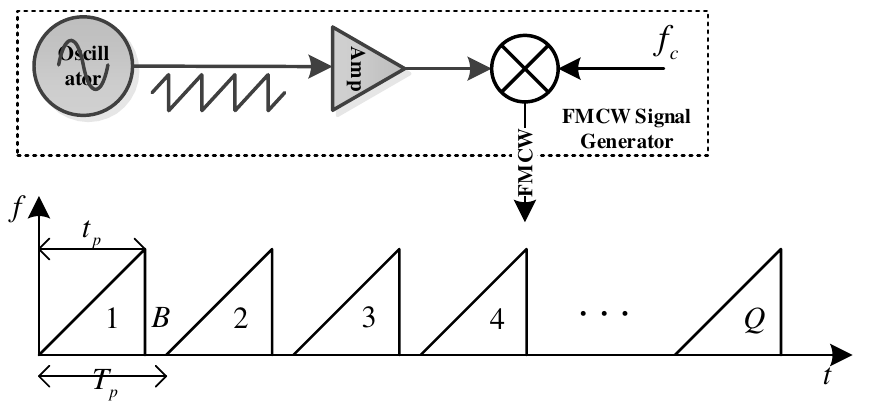}}
    \\
  \subfloat[TDMA scheme with 4 transmitters.\label{fig1b}]{%
        \includegraphics[width=0.8\linewidth]{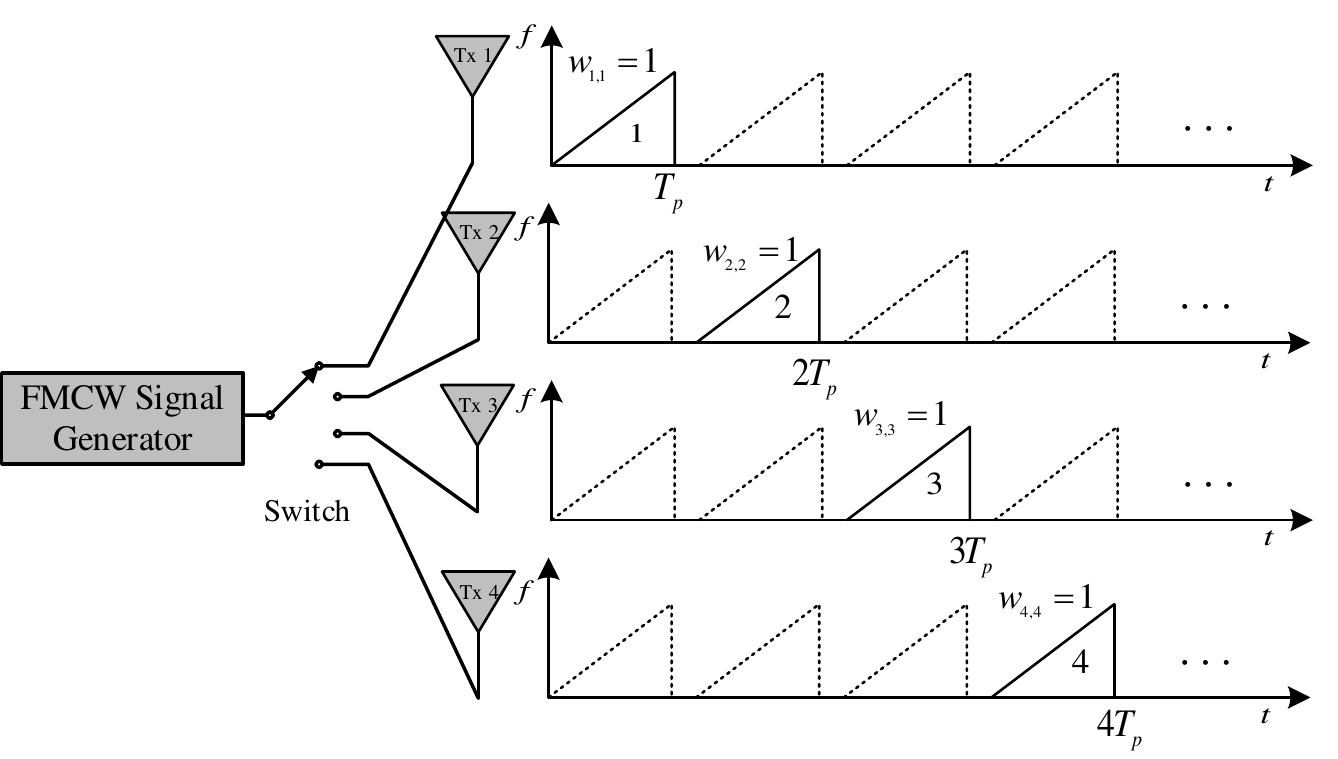}}
  \caption{Automotive MIMO radar transmit system diagram.}
\end{figure}

Consider a single-input single-output (SISO) radar emitting FMCW signal. The baseband waveform can be denoted by ${\bf u}(t) = {\rm rect}\left( \frac{t-T_p/2}{T_p}\right) {e}^{j{\pi}{\beta} t^2}$, where $T_p$ is the pulse repetition interval (PRI), $\beta$ is the chirp rate, $B$ is the bandwidth, and the function ${\rm rect}\left( \frac{t-T_p/2}{T_p}\right)$ represents a pulse centered at $t = T_p/2$ with width $T_p$. Assuming $Q$ pulses in a single coherent processing interval (CPI), the transmitted signal can be given as ${\bf s}(t) = \sum\limits_{q = 0}^{Q - 1} {\bf u}(t-qT_p)e^{j2\pi f_ct}$, where $f_c$ is the carrier frequency (e.g., $ 79~{\rm GHz}$). As shown in Fig.~\ref{fig1a}, the transmitted waveform is repeated $Q$ times to enable the target range and velocity estimation.

Let us consider a target with range $R$ and velocity $v$. After a round-trip time delay $\tau$, the received echo of this target can be regarded as a damped and time-delayed version of the transmitted signal, i.e.,
\begin{equation}\label{eq1}
    \begin{aligned}
        {\bf y}(t) & = \sigma{\bf s}(t-\tau) = \sigma\sum\limits_{q = 0}^{Q - 1} {\bf u}(t-qT_p-\tau)e^{j2\pi f_c(t-\tau)}
   \end{aligned}
\end{equation}
where $\sigma$ is the fading coefficient due to the path and reflection losses. The time delay $\tau$ is given by $\tau  = \tau_0+\tau_d = 2R/c + 2vt/c$ where $c$ is the speed of light. After the carrier frequency downconversion, the baseband signal can be written as ${\bf y}(t)e^{-j2\pi f_ct}$. Then, the signal is conjugately mixed with the source chirp to produce the intermediate-frequency (IF) signal. Inserting $\tau$, the IF signal can be expressed as
\begin{equation}\label{eq2}
  {\bf z}(t) = \sigma\sum\limits_{q = 0}^{Q - 1} {e^{-j2\pi \frac{2(R+vt)}{c}[\beta(t-qT_p)+f_c]}}.
\end{equation}
Note that we split the time index into fast-time and slow-time ($t = p/f_s + qT_p$). Then, the FMCW radar receiver output as a function of these two time indices can be written as
\begin{equation}\label{eq3}
  Z(p,q) = \sigma  \cdot {e^{ - j2\pi (\beta {\tau _0} + {f_d})\frac{p}{{{f_s}}}}} \cdot {e^{ - j2\pi {f_d}q{T_p}}}
\end{equation}
where the factor $\sigma$ absorbs other constant phase factors, the pair $(p,q)$ denotes the sample at the fast-time $p/f_s$ for $q$-th pulse, $f_s$ is the sampling rate, $f_d =2v/\lambda$ is the target Doppler shift, and $\lambda = c/f_c$ is the wavelength.

From \eqref{eq3}, the target range and Doppler information can be estimated by using two-dimensional (2-D) fast Fourier transform (FFT). Typically, the estimation performance can be evaluated by \cite{1}
\begin{equation}\label{eq4}
    \begin{aligned}
       & \Delta_R = \frac{c}{2B},\quad R_{max} = \frac{cf_sT_p}{4B}\\
       & \Delta_v = \frac{\lambda}{2QT_p}, \quad v_{max} = \frac{\lambda}{4T_p}
     \end{aligned}
\end{equation}
where $\Delta_R$ and $\Delta_v$ denote the resolution of range and velocity estimation, respectively, while $R_{max}$ and $v_{max}$ represent the maximum unambiguous range and velocity, respectively.

To enable angle estimation, the received signal should be collected across the spatial dimension via transmit and/or receive array with multiple elements, i.e, the time delay is updated by $\tau = \tau_0+\tau_d+\tau_s$. In the typical case of a narrowband signal in far-field, the extra time delay caused by the phase difference between the $m$-th transmit element and $n$-th receive element is given as $\tau_s = \sin\theta (md_t+nd_r)/c, \; m = 1,\cdots,M, \; n = 1,\cdots,N$, where $\theta$ is the target direction, $d_t$ and $d_r$ are the interelement distances between adjacent elements in the transmit and receive arrays, respectively. In automotive MIMO radar, one of the objectives is to maximize the VA aperture to enhance the angle estimation performance. To achieve it, the key issue is the implementation of orthogonal transmit waveforms. Considering the performance to cost ratio that is of major significance for automotive radar application, the TDMA technique has been recognized as a good choice. Then the received signal at the $n$-th antenna element is given by
\begin{equation}\label{eq5}
   {\bf y}_n(t) = \sigma \sum\limits_{m = 1}^{M} \sum\limits_{q = 0}^{Q - 1} w_{m,q}{\bf u}(t-qT_p-\tau)e^{j2\pi f_c(t-\tau)}
\end{equation}
where $w_{m,q}$ enables the TDMA technique like a switch (see Fig.~\ref{fig1b}).\footnote{It is very difficult in practice to design a high-speed switch that enables the alternative transmitting TDMA technique. We use this concept to just show how the matrix ${\bf W}$ works.} More specifically, we have
\begin{equation}\label{eq6}
  {\bf W} = \underbrace {\left[ {{{\bf{I}}_M},{{\bf{I}}_M},\cdots,{{\bf{I}}_M}} \right]}_{\bar Q}
\end{equation}
where $ {\bar Q} = Q/M$. Updating \eqref{eq3}, the received signal as a function of fast-time, slow-time and spatial direction $\theta$ for the $n$-th receive element can be expressed as
\begin{equation}\label{eq7}
    \begin{aligned}
        {Z}^{(n)}(p,q,\theta){\rm{ }} & = \sigma \sum\limits_{q = 0}^{Q - 1} w_{m,q}\underbrace {{e^{ - j2\pi (\beta {\tau _0} + {f_d})\frac{p}{{{f_s}}}}}}_{Range} \\
        & \underbrace {{e^{ - j2\pi {f_d}q{T_p}}}}_{Doppler}\underbrace {{e^{ - j\frac{{2\pi }}{\lambda }(m{d_t} + n{d_r})\sin \theta }}}_{Angle}.
  \end{aligned}
\end{equation}

Therefore, the target localization and velocity estimation can be jointly conducted using the signal model \eqref{eq7}. The pros and cons of using the TDMA technique are both conspicuous. The signals from distinct transmit elements can be separated easily. The requirement of only one signal generator implies that the automotive MIMO radar can be upgraded from its SIMO counterpart conveniently. On the other hand, the transmit power is reduced while the velocity estimation capability is deteriorated, as the transmit cycle for each antenna is increased $M$ times to ensure only one antenna is active at any given time. If $M$ raises, it can be seen that the TDMA technique applied for automotive MIMO radar either suffers from the loss of $v_{max}$ and potential range migration or is afflicted with the reduced $R_{max}$. These drawbacks can be fatal for automotive radar application.

\section{Proposed Automotive MIMO Radar Transmit Schemes: TB and DDMA Approaches}
Although the alternative transmitting TDMA technique is easy to implement for automotive MIMO radar, its parameter estimation performance loss can be unacceptable in some scenarios. The critical weakness of the TDMA approach is that the available time resource is always confined in a way, and this becomes more evident with the increase of the number of transmit elements. Another problem is the loss of transmit energy. Our goal is to design a new transmit scheme so that the automotive MIMO radar can fully exploit the transmission capability of the transmit array and maintain the parameter estimation ability as much as possible.

\subsection{TB Approach}
 In TB technique \cite{11,12}, the required number of orthogonal waveforms can be reduced to only two regardless of the number of transmit elements. This property is attractive, since it helps to avoid the overwhelming trade-off between the active time for each antenna (slot time) and the overall transmit time duration (frame time). Moreover, the omnidirectional transmit beampattern for the conventional MIMO radar is modified by the TB matrix to generate a relatively fat beampattern that covers only the spatial region of interests. For example, the required azimuthal FoV is approximately $\pm 15^\circ$/$\pm 40^\circ$ for LRR/MRR applications \cite{1}. The TB approach that focuses the transmit energy on a fixed spatial area clearly can surpass the conventional TDMA approach.

Consider an FMCW signal generator using the TDMA transmit scheme to create $K = 2$ orthogonal waveforms. In one single frame containing $Q = K$ time slots, the orthogonal waveforms are given by ${\bf s}_k(t) = {\bf u}(t-kT_p)e^{j2\pi f_ct}, t \in [0, KT_p]$. Using the TB technique, the signal transmitted by each antenna element is replaced by the combination of the waveforms ${\bf s}_k(t)$. Assuming a uniform linear array (ULA) at the transmitter with $M \ge K$ antenna elements, the synthesis problem for the transmit power distribution can be formulated as the following minimax problem \cite{12}
\begin{equation}\label{eq10}
  \mathop {\min }\limits_{\bf D} \mathop {\max }\limits_\theta  \left| {P_i(\theta) - \sum\limits_{k = 1}^K {{\bf{d}}_k^H{\bf{a}}(\theta ){{\bf{a}}^H}(\theta ){{\bf{d}}_k}} } \right|
\end{equation}
where $P_i(\theta)$ is the ideal transmit beampattern, e.g., a rectangular shape ${\rm rect}\left( {\theta}/{\theta_0}\right)$, with transmitted energy focused within the region $\Theta : [-\theta_0,\theta_0]$, ${\bf a}(\theta) \triangleq [1,e^{-j\frac{2\pi}{\lambda}d_t\sin\theta},\cdots,e^{-j\frac{2\pi}{\lambda}(M-1)d_t\sin\theta}]^T$ denotes the transmit steering vector at direction $\theta$, and ${\bf D} \triangleq [{\bf d}_1,{\bf d}_2,\cdots, {\bf d}_K]$ is the TB matrix of size ${M \times K}$.

By solving optimization problem \eqref{eq10}, we can generate the desired TB matrix. However, there are some practical design constrains that need to be considered in addition. These constraints are the following. First, it is necessary to achieve a unit transmit power on each transmit element so that the amplifier would operate in linear area. Moreover, since the waveforms transmitted by different antenna elements are separated in time slots, the transmit power of each waveform should also be identical to enable the beamforming as well as other operations such as DOA estimation at the receiver side. These practical constrains can be expressed as scaling the rows and columns of the TB matrix ${\bf D}$ so that the Euclidean norms of the rows are the same and the Euclidean norms of the columns are the same, or equivalently, as
\begin{equation}\label{eq11}
  \sum\limits_{m{\rm{ = 1}}}^M {{{\left| {d_{m,k}} \right|}^2}}  = {c_1}, \quad  \sum\limits_{k{\rm{ = 1}}}^K {{{\left| {d_{m,k}} \right|}^2}}  = {c_2}
\end{equation}
where $c_1$ and $c_2$ are some constants. Combining \eqref{eq10} and \eqref{eq11}, a non-convex optimization problem can be formulated. Note that the number of waveforms $K$ in automotive MIMO radar scenario is naturally small.\footnote{A typical automotive MIMO radar based on the TDMA technique is equipped with three transmit elements \cite{8}. Thus, it is practically reasonable to assume that $K$ is at the same level.} This fact can be exploited to relax the non-convex optimization problem of TB design. Considering for example $K = 2$, the non-convex optimization problem of TB design can be formulated as
\begin{align}
& \mathop {\min }\limits_{{\bf d}_1,{\bf d}_2} \mathop {\max }\limits_\theta  \left| {P_i(\theta) - \sum\limits_{k = 1}^2 {{\bf{d}}_k^H{\bf{a}}(\theta ){{\bf{a}}^H}(\theta ){{\bf{d}}_k}} } \right|\label{eq12}\\
& s.t.\quad||{\bf d}_1||^2 =  ||{\bf d}_2||^2 = c_1 \tag{\ref{eq12}{a}} \label{eq12a}\\
& \quad \quad  {{{\left| {d_{m,1}} \right|}^2}}+ {{{\left| {d_{m,2}} \right|}^2}}  = {c_2}\tag{\ref{eq12}{b}} \label{eq12b}\\
& \quad \quad  m = 1,\cdots,M. \notag
\end{align}

By exploiting the conjugate symmetry property, i.e., ${\bf J}_M{\bf a}^*(\theta) = e^{j\frac{2\pi}{\lambda}(M+1)d_t\sin\theta}{\bf a}(\theta)$, we have
\begin{equation}\label{eq13}
  e^{-j\frac{2\pi}{\lambda}(M+1)d_t\sin\theta}\left[({\bf J}_M{\bf d}^*_k)^H{\bf a}(\theta)\right] = ({\bf d}_k^H {\bf a}(\theta))^* .
\end{equation}
Consequently, the amplitudes of power distribution function generated by ${\bf d}_k$ and ${\bf J}_M{\bf d}_k^*$ are identical. If we let ${\bf d}_2 = {\bf J}_M{\bf d}_1^*$, constraint \eqref{eq12a} is satisfied automatically. Note that $|{{\bf{d}}_1^H{\bf{a}}(\theta ){{\bf{a}}^H}(\theta ){{\bf{d}}_1}}| = tr\{{\bf d}_1{\bf d}_1^H{\bf a}(\theta){\bf a}^H(\theta)\}$. Then, problem \eqref{eq12} can be reformulated as
\begin{equation}\label{eq14}
    \begin{aligned}
        & \mathop {\min }\limits_{{\bm \Delta},t} \quad t \\
        & s.t.\quad \left |{P_i(\theta)} - 2\cdot tr\left\{{\bf a}(\theta_j){\bf a}^H(\theta_j){\bm \Delta}\right\}\right| \le t, \theta_j \in \Theta\\
        & \quad\quad tr\left\{{\bf \Delta}{\bf \Gamma}^{(m)}\right\} = 1,\quad m = 1,2,\cdots, \left\lceil \frac{ M}{2} \right\rceil
    \end{aligned}
\end{equation}
where ${\bf \Gamma}^{(m)} = diag({\bm \gamma}^{(m)})$, ${\bm \gamma}^{(m)} \in \mathbb{C}^{M}$ is an all-zero vector except that its $m$-th and $(M-m+1)$-th elements being one, ${\bf \Delta} = {\bf d}_1{\bf d}_1^H$ is a rank-one positive semidefinite matrix, and we let the scaling factor $c_2$ to be just 1 for simplicity.

Problem \eqref{eq14} is convex and can be solved by standard semidefinite programming (SDP) solvers. Hence, the corresponding eigenvector ${\bf d}$ is the required beamspace vector and the TB matrix is given by
\begin{equation}\label{eq15}
  {\bf D} = [{\bf d}, {\bf J}_M{\bf d}^*] .
\end{equation}

\begin{remark}:
Denote the optimal solution of problem \eqref{eq14} as ${\bf \Delta}^\star$. If its rank is larger than one, then the randomization techniques \cite{12} can be utilized to reduce it to a rank-one approximation. Although the result \eqref{eq15} is designed for $K = 2$, it can be easily generalized to the case of $K = 4$ by letting ${\bf D} = [{\bf d}, {\bf J}_M{\bf d}^*,{\bf d}^*, {\bf J}_M{\bf d}]$ and the order of column vectors can be arbitrary. The further increase of $K$ would be beyond the purpose of using the TB approach for automotive radar application.
\end{remark}

\subsection{DDMA Approach}
\begin{table*}[!t]
\renewcommand{\arraystretch}{1.3}
\caption{Comparisons of TDMA, DDMA, PMCW and Hadamard Code Waveforms}\label{CodeFamily}
\centering
\begin{tabular}{|c||c||c||c|}
\hline
Code Type & Character & Code Length & Comments\\
\hline
TDMA &  Binary (repeated identity matrix) & flexible (M) & easy implementation, transmit energy loss \\
\hline
DDMA & Arbitrary(\eqref{eq16} and \eqref{eq18}) & flexible (M) & quasi-orthogonality, only for SRR application\\
\hline
PMCW & Binary(e.g., Barker, Gold, APAS) & arbitrary & full transmit capacity, Doppler intolerance\\
\hline
Hadamard code & Binary (repeated Hadamard matrix) & confined & simple structure, Doppler ambiguity \\
\hline
\end{tabular}
\end{table*}

Another transmission scheme for automotive MIMO radar can be developed based on the DDMA approach, which refers to a phase modulation for each transmit element from pulse to pulse. DDMA MIMO radar inherits the property of TDMA MIMO radar, i.e., it can be easily upgraded from its SIMO counterpart with a number of additional phase shifters at the transmitter side. The DDMA approach can fully take advantage of the transmission capabilities of the transmit array as all transmit elements can emit simultaneously at any time. However, the waveform diversity is achieved at the cost of a higher level PRF tolerance. To meet the requirements on the maximum unambiguous velocity $v_{max}$, the system PRI needs to be decreased by $M$ times. Consequently, the DDMA scheme suffers from shorter chirp time and is only suitable for SRR applications.

Consider the signal model of the DDMA technique. Like in \cite{23}, we call ${\bf W} \triangleq [{\bf w}_1,{\bf w}_2,\cdots, {\bf w}_Q]$ as phase modulation matrix. Each row vector of ${\bf W}$ behaves as a phase modulator in Doppler domain, i.e., the zero Doppler in range-Doppler map for one particular transmit element is modulated from zero to a unique Doppler frequency $f_m = {(m-1)f_a}/{M}, \; m = 1, 2, \cdots, M$, where $f_a = {1}/{T_p}$. Hence, the phase modulation vector for the $q$-th pulse is given by
\begin{equation}\label{eq16}
  {\bf w}_q = e^{-j2\pi{\bf f}qT_p}
\end{equation}
where ${\bf f} \triangleq [f_1,f_2,\cdots,f_M]^T$ denotes the pre-designed Doppler shift vector for all transmit elements. The row vector for the $m$-th transmit element is given as ${\bm \omega}_m \triangleq [1, e^{-j2\pi f_mT_p}, \cdots, e^{-j2\pi f_m(Q-1)T_p}]$. The elements of this vector are given by the phase function $e^{-j2\pi f_m t}$ where $t = qT_p$ in slow-time. Comparing to \eqref{eq6}, it can be seen that the DDMA technique shares the same transmitter structure with the TDMA technique. While the elements of ${\bf W}$  in the TDMA technique are binary (either zero or one), the implementation of the DDMA technique generally requires $M$ different phase shifts. Consequently, all transmit elements emit the phase modulated waveforms simultaneously in the DDMA technique. Thus, the transmission capability of the transmit array is fully exploited as compared to the TDMA technique.

To further explain the essence of the DDMA approach, let us consider the signal  received by the $n$-th receive element in DDMA MIMO radar. Rewrite \eqref{eq7} as
\begin{equation}\label{eq17}
\begin{aligned}
  {Z}^{(n)}(p,q,\theta) = \sigma \sum\limits_{q = 0}^{Q - 1} e^{-j2\pi f_mqT_p} {{e^{ - j2\pi (\beta {\tau _0} + {f_d})\frac{p}{{{f_s}}}}}}\\
   {{e^{ - j2\pi {f_d}q{T_p}}}} {{e^{ - j\frac{{2\pi }}{\lambda }(m{d_t} + n{d_r})\sin \theta }}}
\end{aligned}
\end{equation}
and apply the FFT in slow-time. It can be observed then that the phase modulation factor $e^{-j2\pi f_mqT_p}$ shifts the signal transmitted by the $m$-th transmit element with an additional Doppler frequency $f_m$. Note that $f_m$ is uniformly spaced on the entire Doppler spectrum refined by $\left[ -{f_a}/{2}, {f_a}/{2} \right]$, the unambiguous Doppler region for each transmit-receive channel is therefore reduced by $M$ times to its step size, i.e., $\Delta f = f_{m+1}-f_m$.
It is worth noting that after the 2-D FFT is applied on the received signals, a smaller range-Doppler map for each transmit-receive channel can be obtained by cutting off the original range-Doppler map from different Doppler shifts $f_m$.

\begin{figure}
\centerline{\includegraphics[width=0.8 \columnwidth]{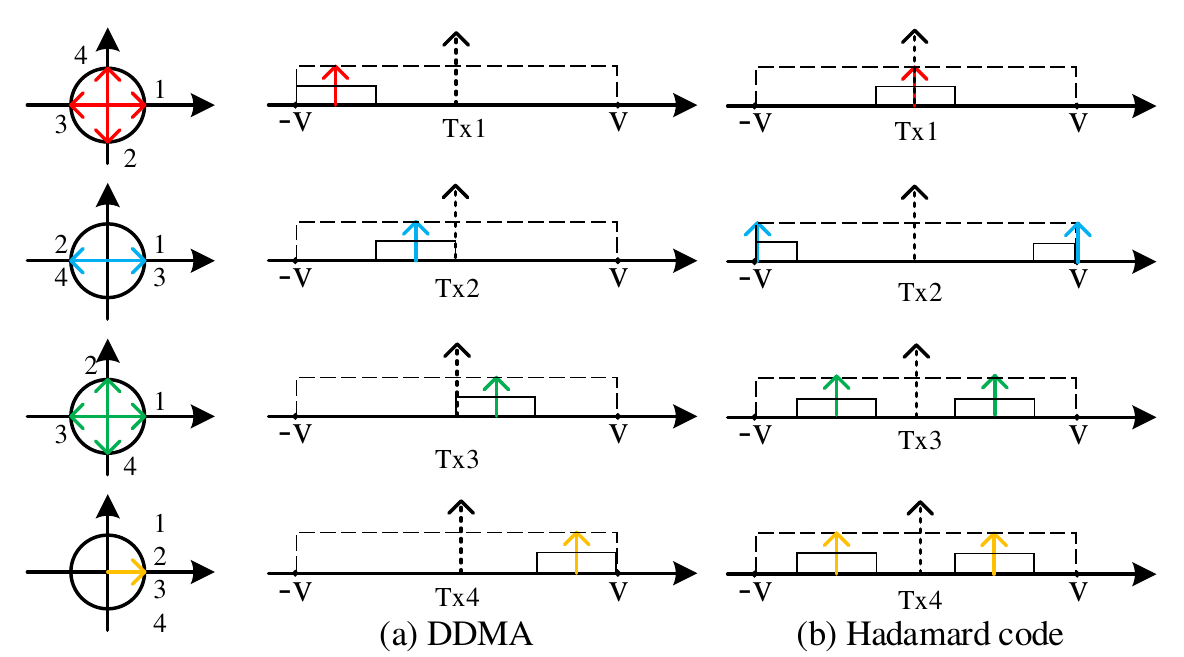}}
\caption{Diagram of Doppler shifts in range-Doppler map for the DDMA approach with 4 transmit elements.}\label{Modulation}
\end{figure}

Summarizing, the waveform diversity can be achieved in Doppler domain for automotive MIMO radar. To keep $v_{max}$ unchanged, the radar PRI must be reduced by $M$ times accordingly. Fig.~\ref{Modulation} demonstrates the Doppler shifts for the DDMA approach with 4 transmit antennas. The signal transmitted by each antenna is modulated by a uniformly increasing phase sequence at different step sizes (e.g., $3\pi/2, \pi, \pi/2,0$).

It can be observed that the rows and columns of ${\bf W}$ for the DDMA approach are both Vandermonde vectors with constant amplitude. Together with the TDMA approach, the DDMA approach can be categorized as a special case of phase modulated continuous wave (PMCW) radar \cite{14,15,16}. The only difference is the structure of the matrix ${\bf W}$, where different classes of codes like Barker, Gold, and almost perfect autocorrelation sequence (APAS) are usually applied. Comparing with them, the DDMA apporach overcomes the weakness of Doppler intolerance that destroys the orthogonality principles. On the other hand, the binary sequences used in PMCW radar have relaxed implementation requirements to hardware while the extra phase $e^{j\phi}$ added in the DDMA approach is arbitrary and demands sufficient accuracy.

There exists another special binary code with almost perfect orthogonality that can be used to design ${\bf W}$, which is Hadamard code \cite{6}. The advantage of Hadamard code is that the matched-filtering of the received signals is very simple, it requires only basic sum and difference operators. The number of transmit elements as well as the code length are confined due to the property of Hadamard matrix. In fact, the Hadamard code waveform is somewhat similar to the DDMA waveform. Taking Fig.~\ref{Modulation} as an example of the Hadamard code and DDMA approach with $M = 4$, we can see that the first two antenna elements with phase step size of $0,\pi$ are unchanged, while the other two antenna elements are modulated by $\cos 2\pi f_m t$ and $\sin 2\pi f_m t$ in the Hadamard code instead of $e^{-j2\pi f_m t}$ in the DDMA approach. This reduces the system complexity, but leads to ambiguous Doppler shifts, as shown in Fig.~\ref{Modulation}(b). The Hadamard code has been previously introduced for automotive MIMO radar (see \cite{25}), where Doppler compensation had to be also considered. Similar to \eqref{eq6}, the Hadamard matrix of size $M \times M$ is repeated $\bar Q$ times to form the phase modulation matrix. Table~\ref{CodeFamily} summarizes the properties of the TDMA, DDMA, PMCW, and Hadamard code waveforms.

\begin{remark}:
  If the waveforms are coded via $f_m = {(m-1)f_a}/{M}$, there is actually no phase modulation for the first transmit element. The corresponding smaller range-Doppler map is just the original range-Doppler map with reduced Doppler spectrum. For some transmit elements with phase modulation $f_m = \pm {f_a}/{2}$, however, the corresponding range-Doppler map may be separated into two parts since the zero Doppler baseline appears at the margin of the original range-Doppler map (see Fig.~\ref{Modulation}(b)). To prevent this and to shift the modulated zero Doppler frequencies of all antenna elements from $-{f_a}/{2}$ to ${f_a}/{2}$ sequentially (see Fig.~\ref{Modulation}(a) and also \cite{24}), we usually let $f_m = 0.5 {f_a} \left( -1 + {(2m-1)}/{M} \right)$. Thus, there is no significant differences between these two coding methods except for the zero Doppler shift locations.
\end{remark}

Next, while the rows of ${\bf W}$ shift the range-Doppler plain, the columns of ${\bf W}$ lead to virtual transmit beamforming. Note that there is in fact only one waveform in the DDMA technique\footnote{In the DDMA approach, all transmitted waveforms are generated using the same FMCW signal.}, and the column vector ${\bf w}_q$ acts as a beamforming vector in every single pulse. The transmitted energy is focused on a particular direction and this particular direction varies from pulse to pulse periodically. For a ULA with $M$ transmit elements, if \eqref{eq16} is applied, the virtual transmit beamforming has $M$ distinct directions $\theta_{\rm v}$  and the phase modulation matrix is given by
\begin{equation}\label{eq18}
\begin{aligned}
  &\sin\theta_{\rm v} = \frac{2q^{\star}}{M}, \quad q^{\star} \in {\mathbb{Z}}\\
  &{\bf W} = \underbrace {\left[ {{\bf \Omega}, {\bf \Omega},\cdots,{\bf \Omega}} \right]}_{\bar Q}\\
  \end{aligned}
\end{equation}
where $q^{\star} \in \left( - \rm{M}/{2}, \rm{M}/{2} \right]$ gives $M$ integers, the matrix ${\bf \Omega} \in \mathbb{C}^{M \times M}$ contains $M$  beamforming vectors corresponding to distinct directions $\theta_{\rm v}$. See Appendix~\ref{appendixA} for more details.

These beamforming vectors (i.e., columns of ${\bf W}$) together (i.e., whole ${\bf W}$) define the DFT-based TB in slow-time \cite{30}. It has the same application features as the previous described TB technique. The difference is that the previously described TB technique is conducted in fast-time while the DFT-based TB generated by the DDMA approach acts in slow-time. From this point of view, the proposed two transmit schemes for automotive MIMO radar essentially yield the TB designs in two different ways. This observation immediately leads to the following questions: {\it is it possible to focus the transmit energy on a fixed region using DDMA approach? If so, could we design the phase modulation matrix, i.e., the TB matrix in slow-time, in a more efficient way?}

\section{TB DDMA Approach with Doppler Ambiguity Mitigation}
For automotive MIMO radar based on the DDMA approach, we have demonstrated that the waveform diversity can be achieved, but it is achieved at the cost of high PRF that reduces the maximum unambiguous range $R_{max}$. If we wish to keep $R_{max}$ (i.e. keep the radar PRF), the problem of Doppler ambiguity caused by the DDMA waveform needs to be solved. Furthermore, we can introduce the TB technique to focus the transmitted chirp signal within the region of interest in order to increase the power that can be reflected by the targets. It should be stressed again that the Doppler shift on each transmit element and the virtual transmit beamforming from pulse to pulse occur simultaneously in the DDMA approach. In this section, we further develop a new TB DDMA approach with Doppler ambiguity mitigation.

\subsection{Doppler Ambiguity Mitigation}
In the above described DDMA approach, the entire Doppler spectrum is uniformly divided into $M$ pieces. We usually increase the radar PRF to ensure that the maximum unambiguous velocity $v_{max}$ is unchanged after match-filtering in slow-time. However, the declined transmitted energy limits the applicability of the DDMA approach. To tackle this problem, a Doppler ambiguity mitigation method is proposed here using a phase modulation matrix with empty Doppler spectrum, where the number of rows is increased to $M_{\rm v} > M$.

Without loss of generality, let us use \eqref{eq16} to generate a phase modulation matrix of size $M_{\rm v}  \times Q$ and choose the first $M$ rows of this phase modulation matrix to implement the DDMA technique for automotive MIMO radar. Then, for the received signal \eqref{eq17} after applying 2-D FFT, there should be $M$ peaks at the slice of the range-Doppler map from the target range cell. The distance between two adjacent peaks denotes a normalized Doppler shift $1/M_{\rm v}$. Different from the case when $M_{\rm v} = M$, only part of the Doppler spectrum ($M/M_{\rm v}$) is used. Note that the default Doppler shifts on transmit elements are given as a prior information, the normalized maximum unambiguous Doppler shift can be recovered from $\pm 1/2M_{\rm v}$ to $\pm 1/2$ and hence, the Doppler ambiguity mitigation can be performed.

First, let us define the range-Doppler map ${\bf Z}^{(n)} \in \mathbb{C}^{P \times Q}$ as a 2-D FFT of the received signal from the $n$-th receive element. Then, a threshold $T$ is adaptively computed \cite{27,28} via the noise level of $z^{(n)}_{p,q}$ to conduct the detection process. The range-Doppler map can be transformed into a binary matrix ${\bf G}^{(n)} \in \mathbb{R}^{P \times Q}$, with elements given by
\begin{equation}\label{eq19}
g^{(n)}_{p,q} { = } \left\{ \begin{array}{l}
{\rm{1}}, \quad z^{(n)}_{p,q} \ge T\\
{\rm{0}}, \quad {\rm otherwise}
\end{array} \right.
\end{equation}
where 1 indicates that a target is detected while 0 means that no target exists. Then \eqref{eq19} is applied for the signal matrices of all receive elements, and ${\bf{G}}$ is found as
\begin{equation}\label{eq20}
  {\bf{G}} = \bigcap\limits_{n = 1}^N {{{\bf{G}}^{(n)}}} .
\end{equation}
For each row of ${\bf G} \triangleq [{\bf g}_1,{\bf g}_2,\cdots, {\bf g}_P]^T$ (each range cell), if there is a target, a sequence of $M$ 1's with a fixed interval $ Q/M_{\rm v}$ should be observed due to the phase modulation in slow-time.
Then the following function, called hereafter as the test function, can be used to detect the sequence
\begin{equation}\label{eq21}
  H(q) = \left\{ \begin{array}{l}
1,\quad {\rm if } \quad {{\bf{g}}_p}(q) = {\bm{\gamma }}\\
0,\quad {\rm otherwise}
\end{array} \right.
\end{equation}
where ${\bm {\gamma}} \triangleq {\left[ {{\bm{\gamma }}_0^T,\cdots,{\bm{\gamma }}_0^T} \right]^T}$ contains $M$ repeated vectors, ${\bm {\gamma}}_0 \in \mathbb{R}^{Q/M_{\rm v}}$ is a vector with first element being 1 and others being 0, and ${\bf g}(q)$ is a vector of length $QM/M_{\rm v}$ starting from the $q$-th element. Note that FFT in the Doppler domain is cyclic, and the truncated vector ${\bf g}(q)$ also obeys the same rule, i.e., if $q \ge Q - Q M / M_{\rm v}+1$, the off-index elements are selected from the beginning of ${\bf g}(q)$ accordingly.

Then the first $H(q) = 1$ in the $p$-th range cell implies that the element $z^{(n)}_{p,q}$ corresponds to the peak of a target with extra Doppler shift $f_1$. Denote the observed Doppler shift at the $q$-th Doppler cell as $f_{\rm ob}$. The real target Doppler shift is, therefore, given as
\begin{equation}\label{eq22}
  \hat {f_{\rm d}} {\rm{ = }}\psi ({f_{\rm ob}} - {f_1}) = \left\{ \begin{array}{l}
{f_{\rm ob}} - {f_1} - 1,\quad {f_{\rm ob}} - {f_1} > 0.5\\
{f_{\rm ob}} - {f_1},\quad \left| {{f_{\rm ob}} - {f_1}} \right| \le 0.5\\
{f_{\rm ob}} - {f_1} + 1,\quad {f_{\rm ob}} - {f_1} <  - 0.5
\end{array} \right.
\end{equation}
where $\hat {f_{\rm d}}, f_{\rm ob}$ and $f_1$ are normalized. Using the function $\psi (\cdot)$, called hereafter as the thresholding function, we can resolve now targets with a wider Doppler shift range, as compared to the case of the conventional DDMA technique with limited resolution ability $1/2M_{\rm v}$. The improvement comes from the empty Doppler spectrum, which connects different Doppler shifts and the transmit elements accordingly and, therefore, mitigates the Doppler ambiguity. In other words, by increasing the number of virtual transmit elements and using a properly selected part of the phase modulation matrix, the maximum unambiguous velocity $v_{max}$ can be kept without decreasing the radar PRI. The transmitted energy and the maximum unambiguous range $R_{max}$ are hence maintained. It is clear that $M_{\rm v} = M+1$ is sufficient to conduct the proposed Doppler ambiguity mitigation method for the DDMA approach. However, a larger virtual number of transmit elements may be useful in another way, which will be explained in the next section.

\begin{remark}:
  The test function \eqref{eq21} may lead to some false alarms. For example, consider two targets at the same range cell with different velocities. If the velocity is accidentally mixed with the margins of each sub range-Doppler map (i.e., the normalized Doppler shift is integer multiples of $1/2M_{\rm v}$), more than $M$ continuous vectors ${\bm \gamma}_0$ can be identified in ${\bf g}_p$. Then, the test function \eqref{eq21} can return more than two detected targets, where only the first and the last one are true. The others are false alarms. Nevertheless, the false alarm is a negligible error, which barely has influence on the decision center of the automobiles, and the test function works in most scenarios.
\end{remark}

\subsection{DFT-based TB Design}
Note that in the TDMA approach, the TB matrix is designed to focus the transmitted energy using only few waveforms and it requires some optimization tools. To compensate the transmit power loss caused by shorted chirp time, it is also possible to generalize the TB technique to the DDMA approach. By exploiting the virtual transmit beamforming property, a super efficient TB design method that can simultaneously form a DFT-based TB and achieve waveform diversity is proposed next. This approach is named then as the TB DDMA approach.

Note that the transmit array emits the signal to a spatial direction from pulse to pulse in the conventional DDMA approach. A straightforward method to focus the transmitted energy is to consciously select the pulses with particular beam directions from the entire CPI. For example, we can choose only first two pulses in every $M$ pulses to form a TB that consists of two narrow beams. The so generated TB is similar to that in the beamspace root-MUSIC method \cite{26}. However, the pulse selection declines the radar sampling rate in slow-time by $M$ times \cite{23}, which reduces the number of efficient pulses for each antenna element and destroys the one-to-one mapping relationship between the transmit elements and the pre-designed Doppler shifts because of the imbalance of the number of distinct beamforming vectors and the number of distributed Doppler regions. Another problem is that the number of available beam directions is limited by the number of transmit elements. The TB design in slow-time is thus heavily constrained.

To conduct the TB DDMA method, the number of transmit elements and the number of pulses need to be increased virtually. The required TB matrix for the TB DDMA approach can be selected from the virtual phase modulation matrix of larger size. In particular, the spatial region of interest can be divided into $M$ subregions with a fixed spatial interval $\Delta_{\rm v}$, where $M$ is the number of real antenna elements. Note that the beam scanning step size is restricted by $\Delta_{\rm v} = 2/M_{\rm v}$ due to virtual transmit beamforming. The number of virtual transmit elements can therefore be computed. By exploiting the periodical mapping relationship between the beam directions and the columns of the phase modulation matrix, the indices of pulses impinging on the specific region can be determined. The number of virtual pulses $Q_{\rm v} = {\bar Q }M_{\rm v}$ is increased to ensure the decimation of $Q$ pulses, where $Q = {\bar Q} M$ is the number of real pulses. The virtual phase modulation matrix ${\bf W}^{({\rm v})}$ is of size $M_{\rm v} \times Q_{\rm v}$, while the $M \times Q$ elements of the TB matrix in the TB DDMA approach are chosen from ${\bf W}^{({\rm v})}$.

To further illustrate the proposed method, consider again the example of $M = 4$. The original beam directions are $\sin\theta_{\rm v} = \{-0.5,0,0.5,1\}$ with a relatively large step size~0.5. The goal is to design a TB that covers the region $\theta = \{\theta| \sin\theta \in [0,0.3]\}$ with $Q = 256$ pulses and also achieves waveform orthogonality like the conventional DDMA approach. The required step size should be $\Delta_{\rm v} = 0.1$, which means that there are $M_{\rm v} = 20$ virtual transmit elements and $Q_{\rm v} = 1280$ virtual pulses. The virtual phase modulation matrix ${\bf M}^{({\rm v})}$ is given by \eqref{eq16}. The virtual beam directions in this case are $\sin\theta_{\rm v} = 0, 0.1, 0.2, 0.3$. Hence, the first 4 pulses should be selected in every $M_{\rm v}$ pulses. This decimation operator is repeated $\bar Q = 64$ times until we collect the additional phases for $Q = 256$ real pulses. Then, the TB matrix is a submatrix of the matrix with any 4 continuous rows (e.g., the first 4 rows.). Or equivalently, we collect the beamforming vectors of a ULA with $M_{\rm v}$ elements that point at $\sin\theta_{\rm v} = 0, 0.1, 0.2, 0.3$ to form a steering matrix. A submatrix with any 4 continuous rows of the steering matrix is then built. The desired TB matrix is the combination of the $\bar Q = 64$ replica of this constructed submatrix. The designed TB matrix yields a DFT-based TB in slow-time, whose columns correspond to $M$ different beamforming vectors. In Appendix~\ref{appendixB}, we illustrate the dependence of the value of $f_m$ on the pulse index for the conventional DDMA approach and the TB DDMA approach.

The construction of the TB matrix for the TB DDMA approach requires no computation. Indeed, only selection and replication of the beamforming vectors need to be performed. Moreover, the TB matrix keeps the phase modulation property for each transmit element that allows to achieve waveform diversity in Doppler domain\footnote{The normalized unambiguous Doppler region for each transmit antenna is $1/M$ instead of $1/{M_{\rm v}}$, since the $M_{\rm v}$ times decimation in slow-time enlarges the Doppler spectrum for each antenna element.}, while the beamforming vectors are optimized to focus the transmitted energy within a fixed spatial region instead of the entire spatial domain. The proposed TB DDMA approach can fully take advantage of the TB technique, and achieves the waveform diversity in Doppler domain in the same way as the conventional DDMA approach. Note that only phase modulation is used, and the DOA estimation method such as ESPRIT can be conducted straightforwardly after matched-filtering at the receiver side.

If $M_{\rm v}$ is chosen properly, it is also worth noting that the TB DDMA approach can be conducted using the phase modulation matrix with empty Doppler spectrum. After mitigating the Doppler ambiguity, the required TB matrix can actually be selected from the columns of the phase modulation matrix with empty Doppler spectrum. In other words, the received signal using the TB DDMA approach can be directly obtained from its counterpart with empty Doppler spectrum via pulse selection. Since different pulse selection schemes generate different TBs, the TB design can be achieved at the receiver side flexibly.

\section{Simulation Results}
In this section, several simulation examples are performed to demonstrate the feasibility of automotive MIMO radar using the TB DDMA approach. Throughout the simulations, an FMCW signal with bandwidth $B = 300~{\rm MHz}$ and time duration $t_p = 1.5~{\rm us}$ is used. The radar carrier frequency is $79~{\rm GHz}$. During a single CPI, we collect $Q = 512$ chirp signals. The automotive MIMO radar has $M = 8$ transmit elements and $N = 12$ receive elements. Both the transmit and receive antenna arrays are ULA's, and the interelement space between adjacent elements in the transmit and receive arrays are $d_t = \lambda/2$ and $d_r = Md_t$, respectively. The arrays are collocated. The noise is considered to be Gaussian white noise. We assume that there are $L = 3$ targets placed at $50 ~{\rm m},100 ~{\rm m}$ and $150 ~{\rm m}$ with normalized Doppler shifts $0, 1/32$ and $-1/32$ (corresponding to velocities $0 ~{\rm m/s},40~{\rm m/s}$ and $-40~{\rm m/s}$), respectively. The radar cross sections (RCSs) of targets obey the Swerling I model. Note that the signal processing gain is quite high in our settings, the input SNR for different targets is chosen as $-10~{\rm dB}$. Hence, the received signal in a single CPI can be modelled by \eqref{eq7}. The range-Doppler map after 2-D FFT in fast-time and slow-time shows the parameter estimation ability of the waveform directly. The conventional DDMA approach is performed for comparison. The conventional TDMA scheme is also introduced. In the TDMA scheme, the transmit interval of each transmit element is $M t_p$ because of the alternative/repetitive transmission.

\subsection{Example~1: Conventional TDMA and DDMA Approaches}
\begin{figure}
    \centering
  \subfloat[TDMA.\label{fig2a}]{%
       \includegraphics[width=0.5\linewidth]{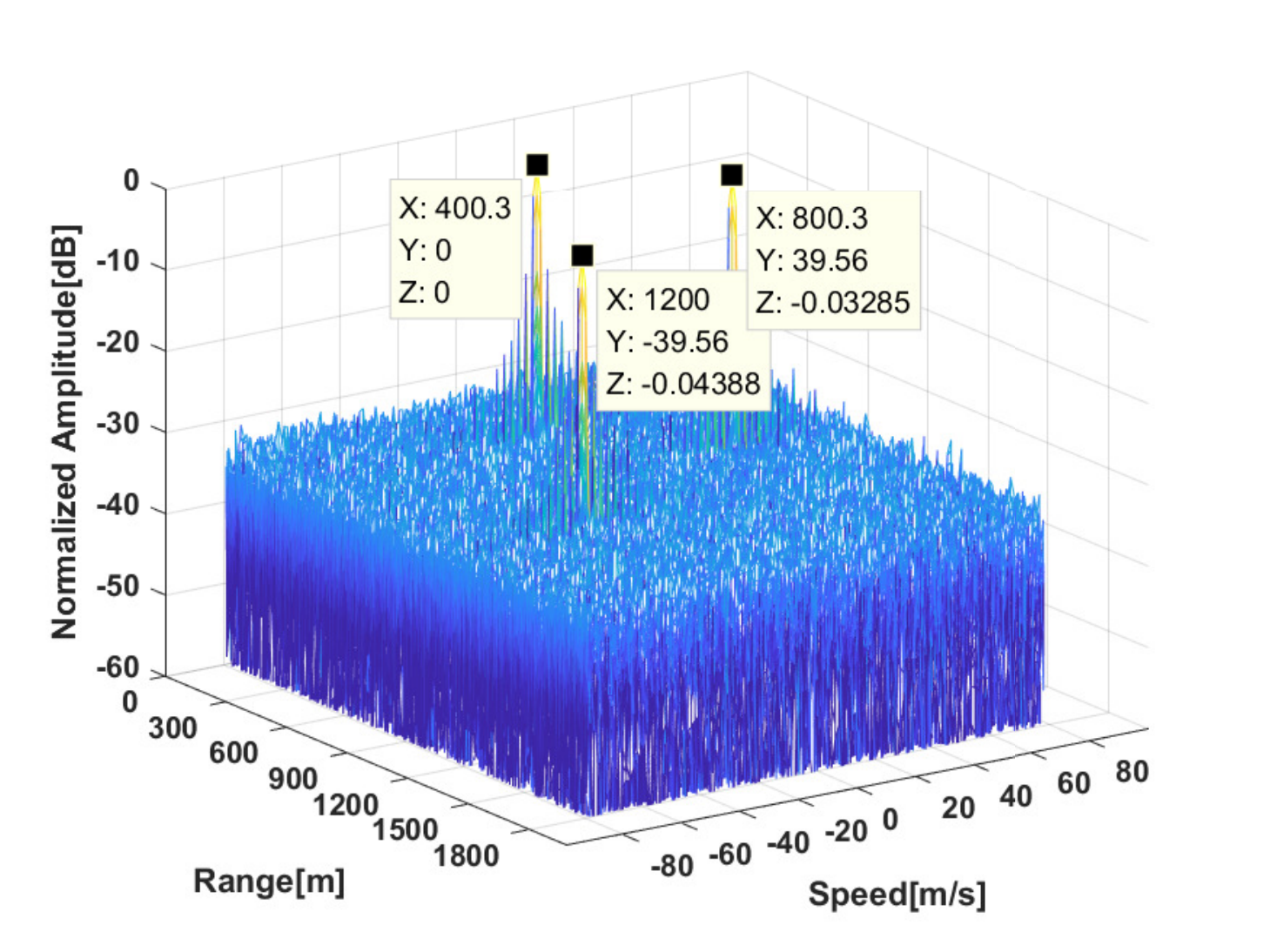}}
    \hfill
  \subfloat[TDMA Slices.\label{fig2b}]{%
        \includegraphics[width=0.5\linewidth]{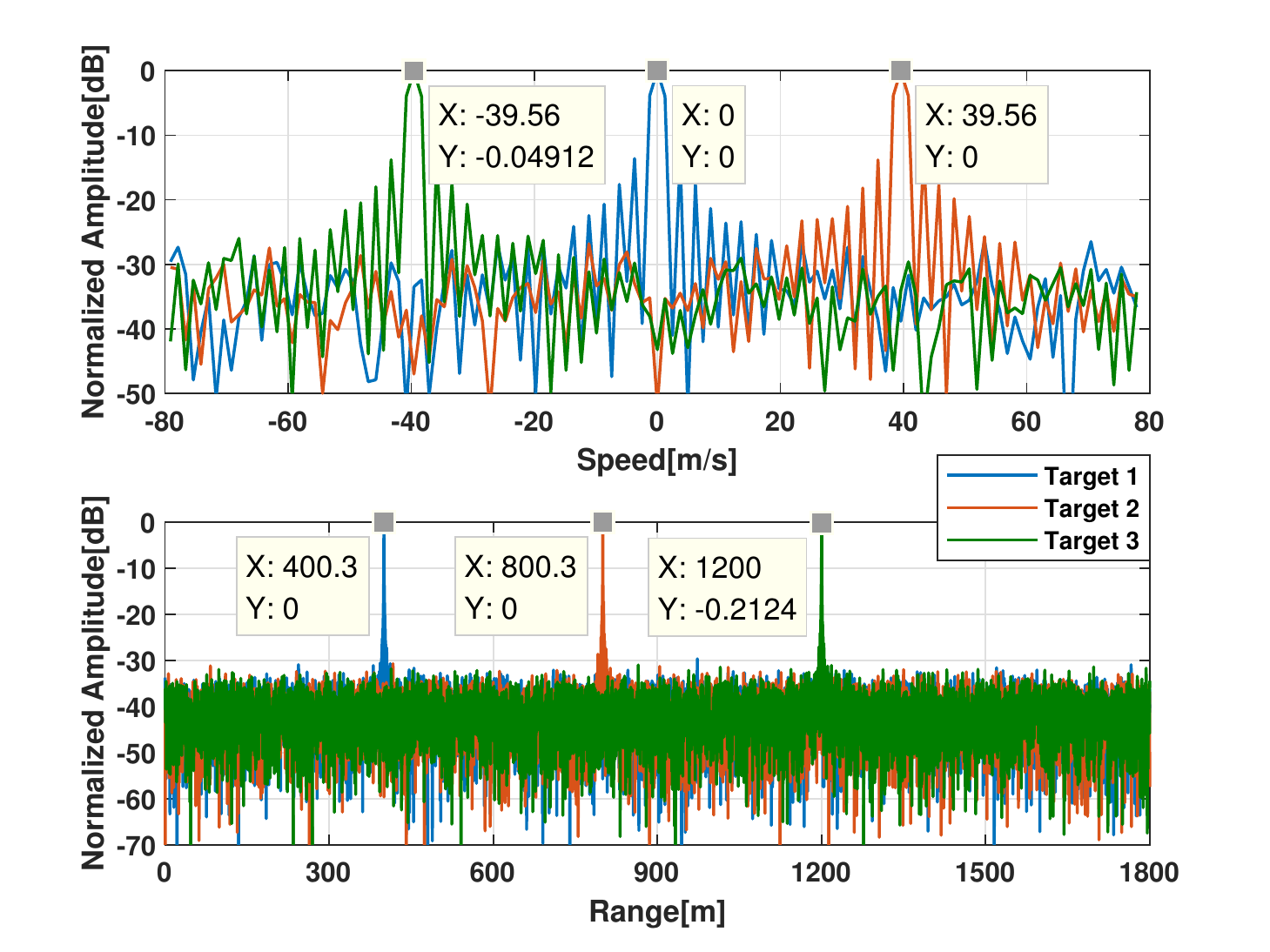}}
  \\
    \subfloat[DDMA.\label{fig2c}]{%
       \includegraphics[width=0.5\linewidth]{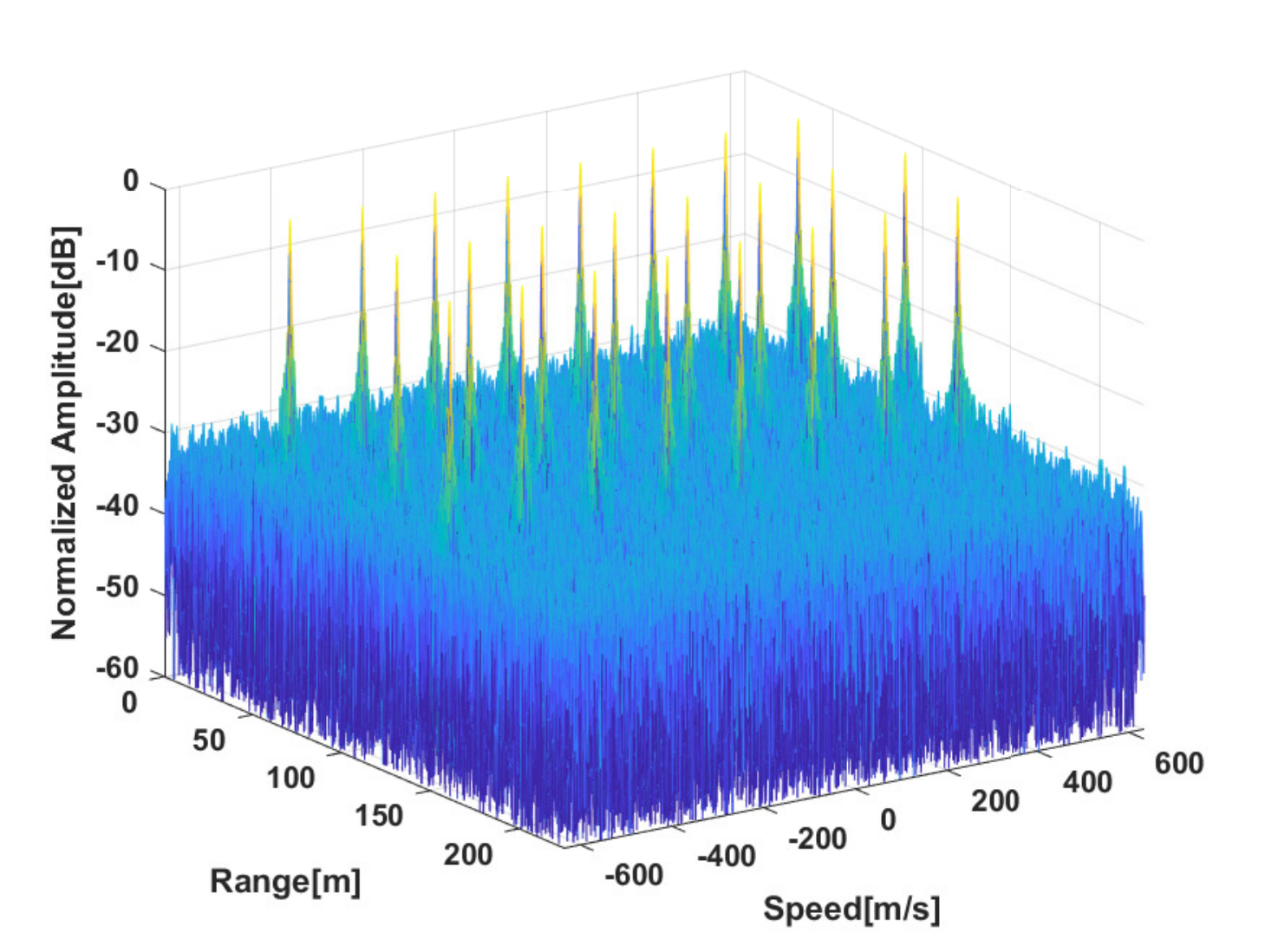}}
       \hfill
         \subfloat[DDMA Slices at target range cells.\label{fig2d}]{%
       \includegraphics[width=0.5\linewidth]{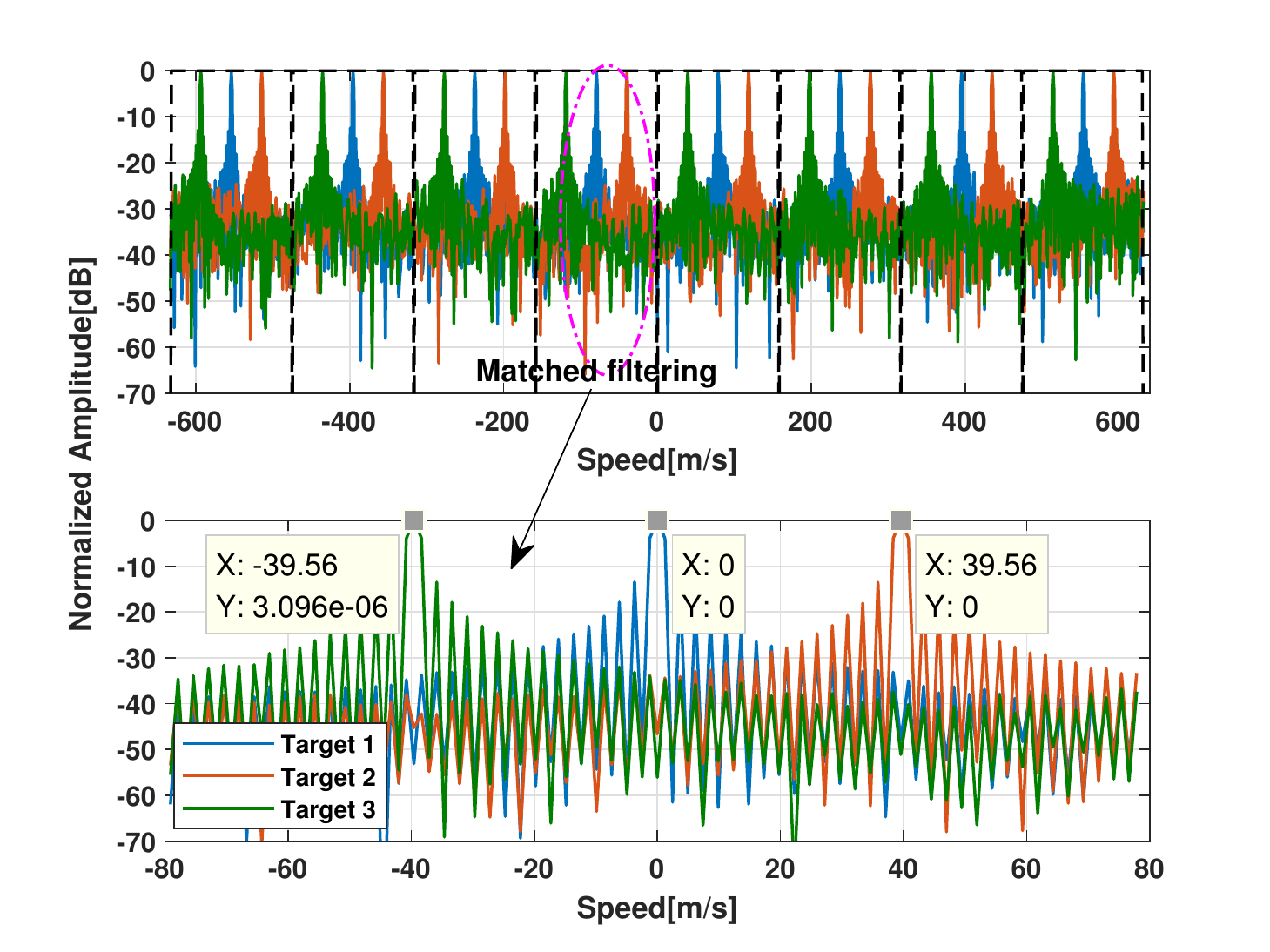}}
  \caption{Range-Doppler map and its slices, three targets.}\label{fig2}
\end{figure}

The TDMA scheme can be fulfilled by using \eqref{eq6}, and the maximum detectable range is increased by $M$ times. Therefore, we let the three targets to appear at $400~{\rm m},800 ~{\rm m}$ and $1200 ~{\rm m}$, accordingly. After 2-D FFT, the range-Doppler map of the received signal in one single receive element is shown in Fig.~\ref{fig2a}, where $R_{max} = 1800 ~{\rm m}$ and $v_{max} = \pm 80~{\rm m/s}$. Fig.~\ref{fig2b} demonstrates the slices of all three targets at the range and Doppler cells. Note that only one transmit element is active. Therefore, the transmitted energy in the TDMA scheme is quite limited that can cause detection problems.

The conventional DDMA approach fully exploits the transmission capability of the transmit array, where the matrix ${\bf W}$ is replaced by \eqref{eq18}. The range-Doppler map of the received signal in one single receive element is shown in Fig.~\ref{fig2c}, where the maximum unambiguous range is only $R_{max} = 225 ~{\rm m}$. Each target has $M$ peaks with identical range cell but different Doppler shifts due to the distinct phase modulation in slow-time. Although the maximum detectable velocity is large (about $\pm 640 ~{\rm m/s}$ ), the Doppler spectrum needs to be divided uniformly into $M$ pieces for each transmit-receive channel to enable the waveform orthogonality. After matched-filtering in slow-time, the maximum unambiguous velocity for each transmit-receive channel is actually the same as that in Fig.~\ref{fig2a} ($v_{max} = \pm 80~{\rm m/s}$). This can be seen in Fig.~\ref{fig2d}.

\subsection{Example~2:  DDMA Approach with Empty Doppler Spectrum}
\begin{figure*}
    \centering
  \subfloat[Range-Doppler map.\label{fig3a}]{%
       \includegraphics[width=0.33\linewidth]{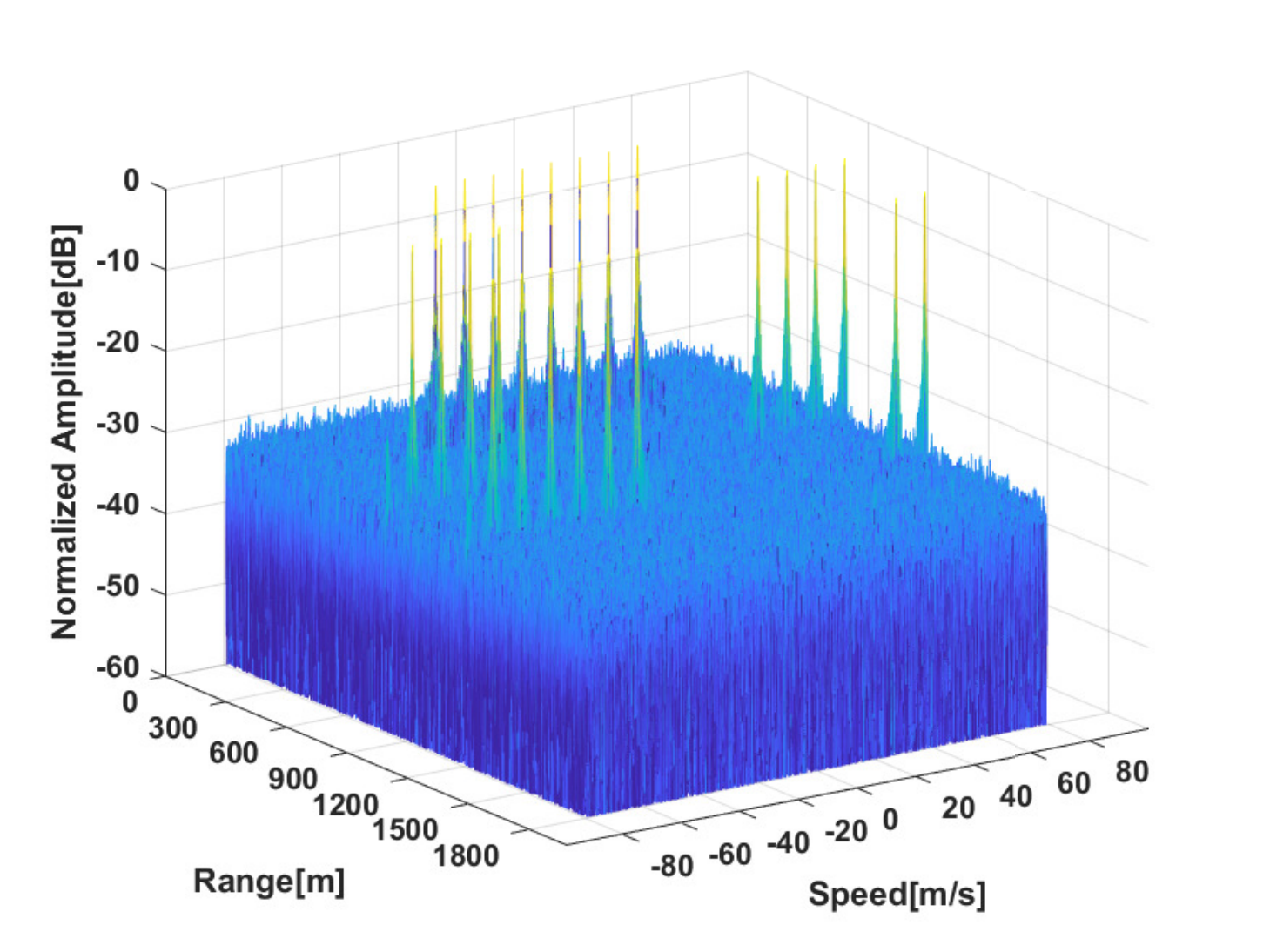}}
    \hfill
  \subfloat[Binary matrix.\label{fig3b}]{%
        \includegraphics[width=0.33\linewidth]{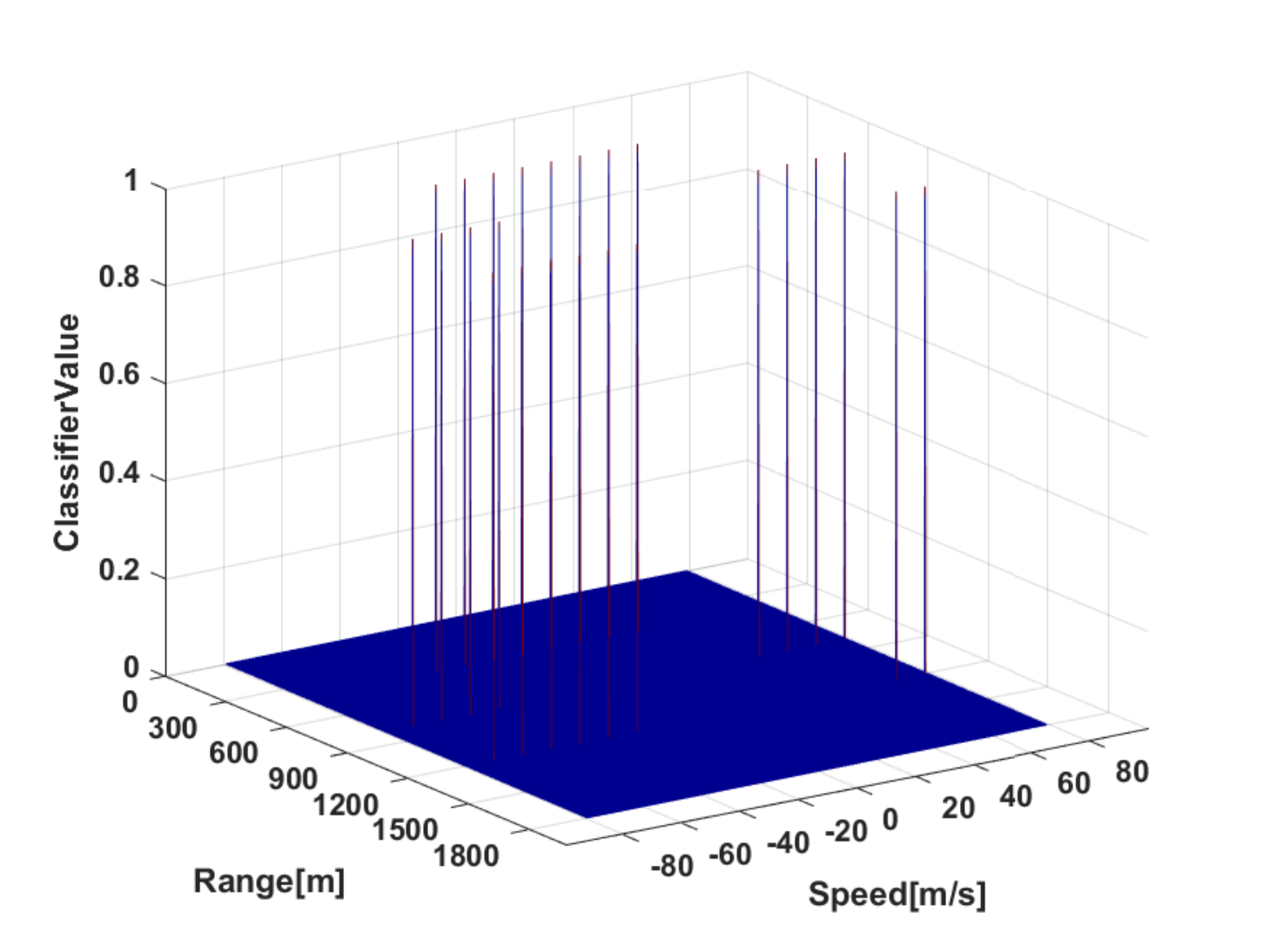}}
    \hfill
  \subfloat[Doppler slices before/after Doppler ambiguity mitigation.\label{fig3c}]{%
        \includegraphics[width=0.33\linewidth]{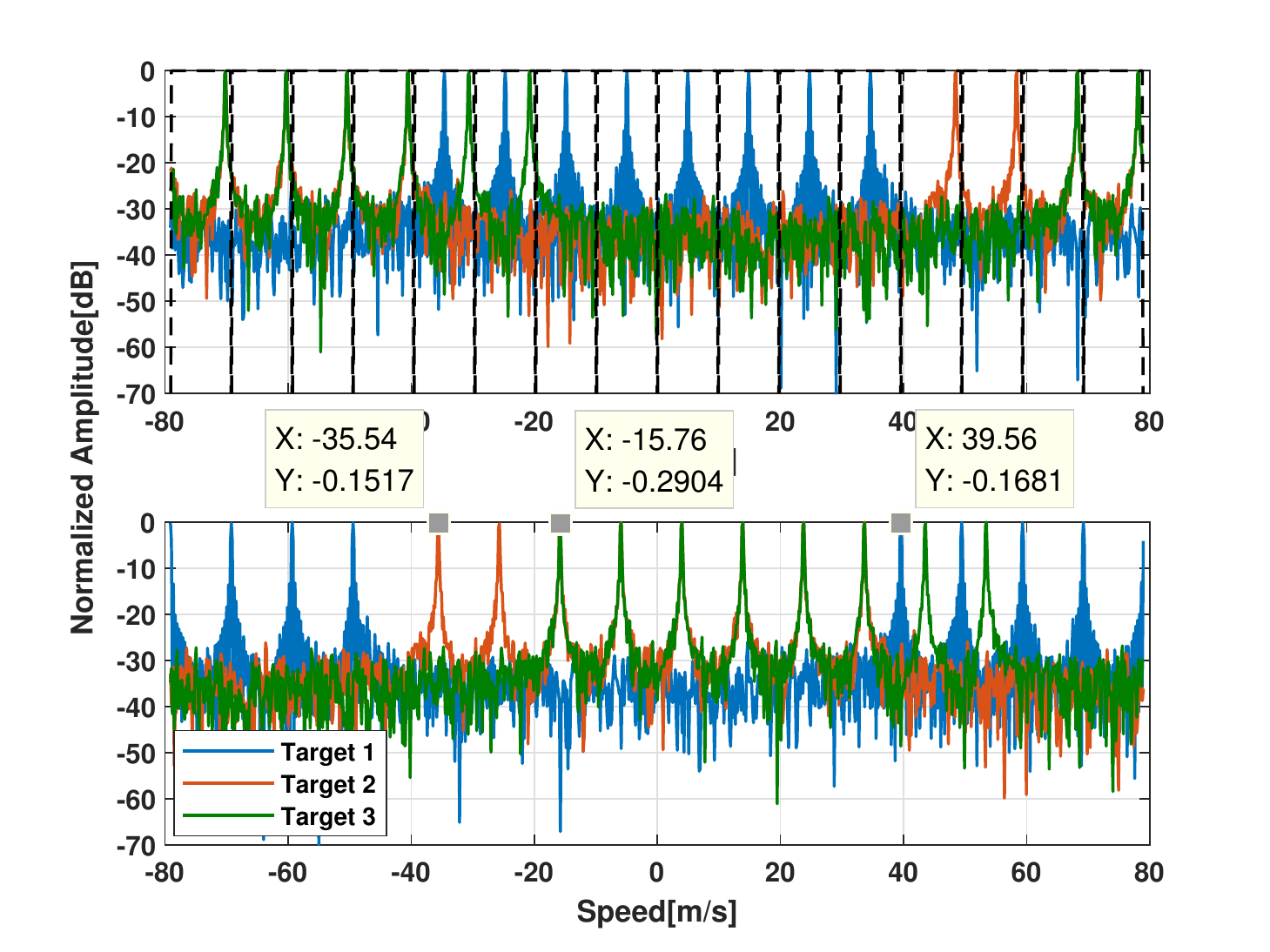}}
\caption{Signal processing results for DDMA with empty Doppler spectrum, three targets.}
\label{fig3}
\end{figure*}

\begin{figure*}
    \centering
  \subfloat[Range-Doppler map.\label{fig4a}]{%
       \includegraphics[width=0.33\linewidth]{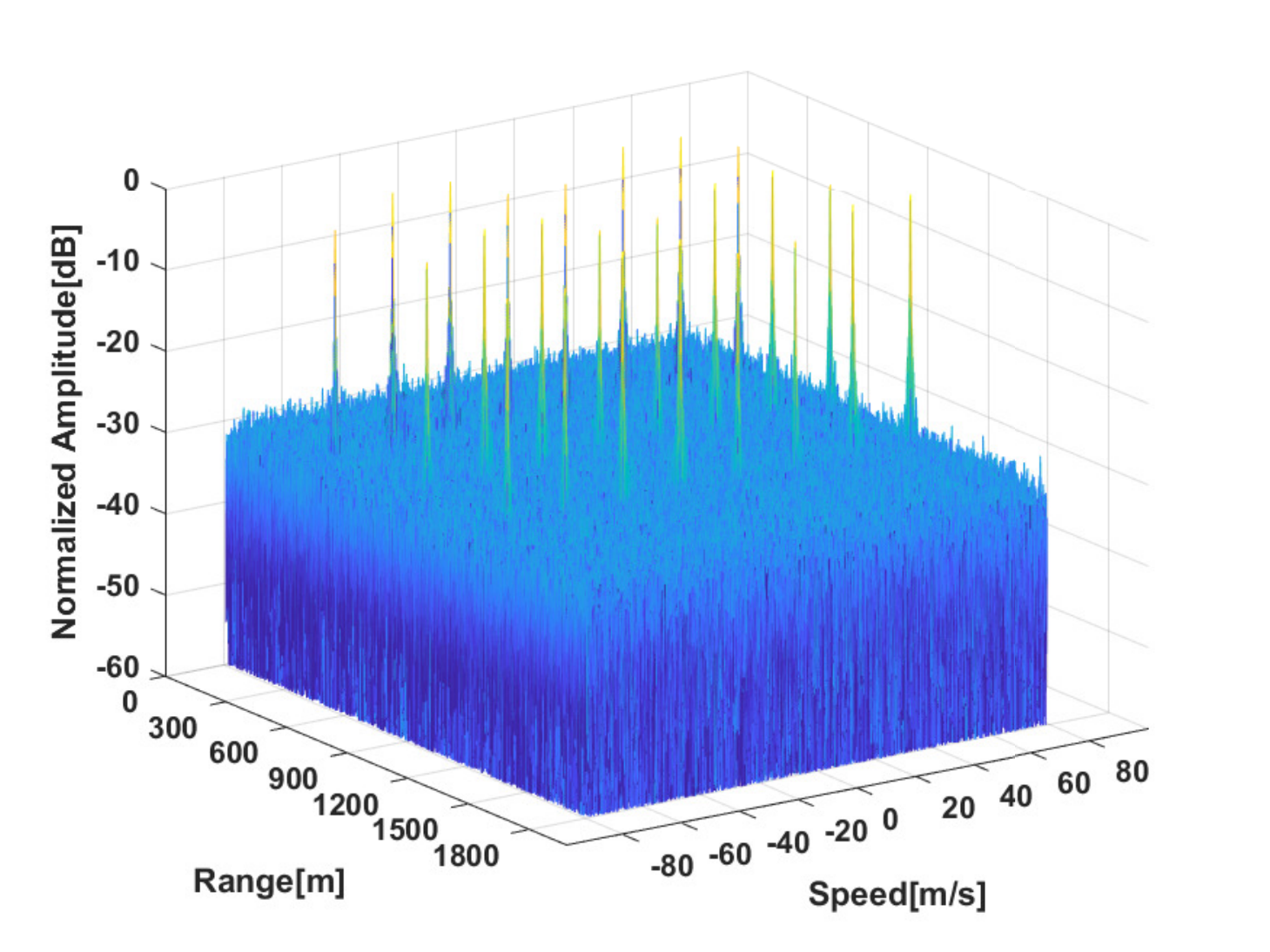}}
    \hfill
  \subfloat[Binary matrix.\label{fig4b}]{%
        \includegraphics[width=0.33\linewidth]{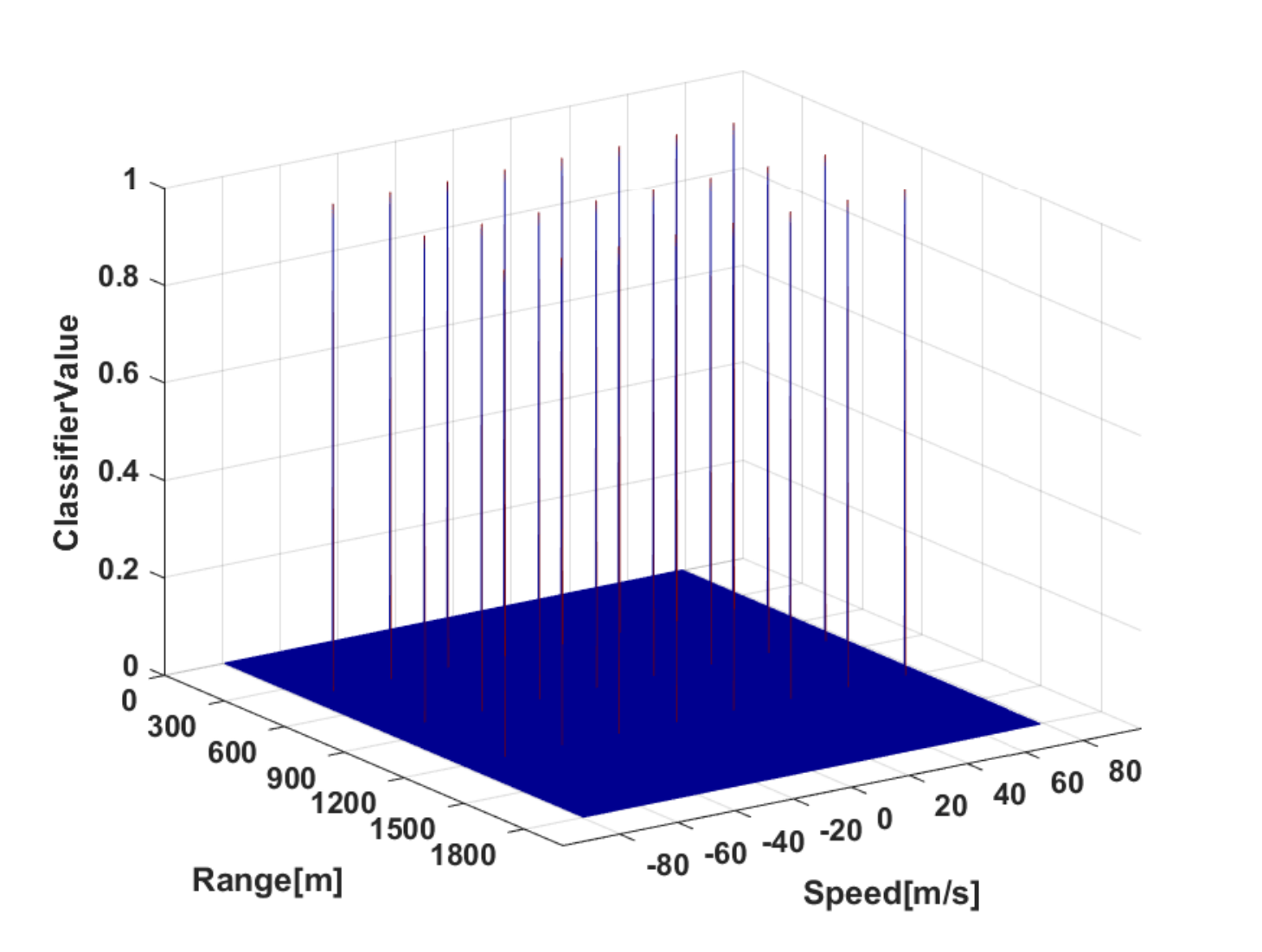}}
    \hfill
  \subfloat[Doppler slices at target range cells.\label{fig4c}]{%
        \includegraphics[width=0.33\linewidth]{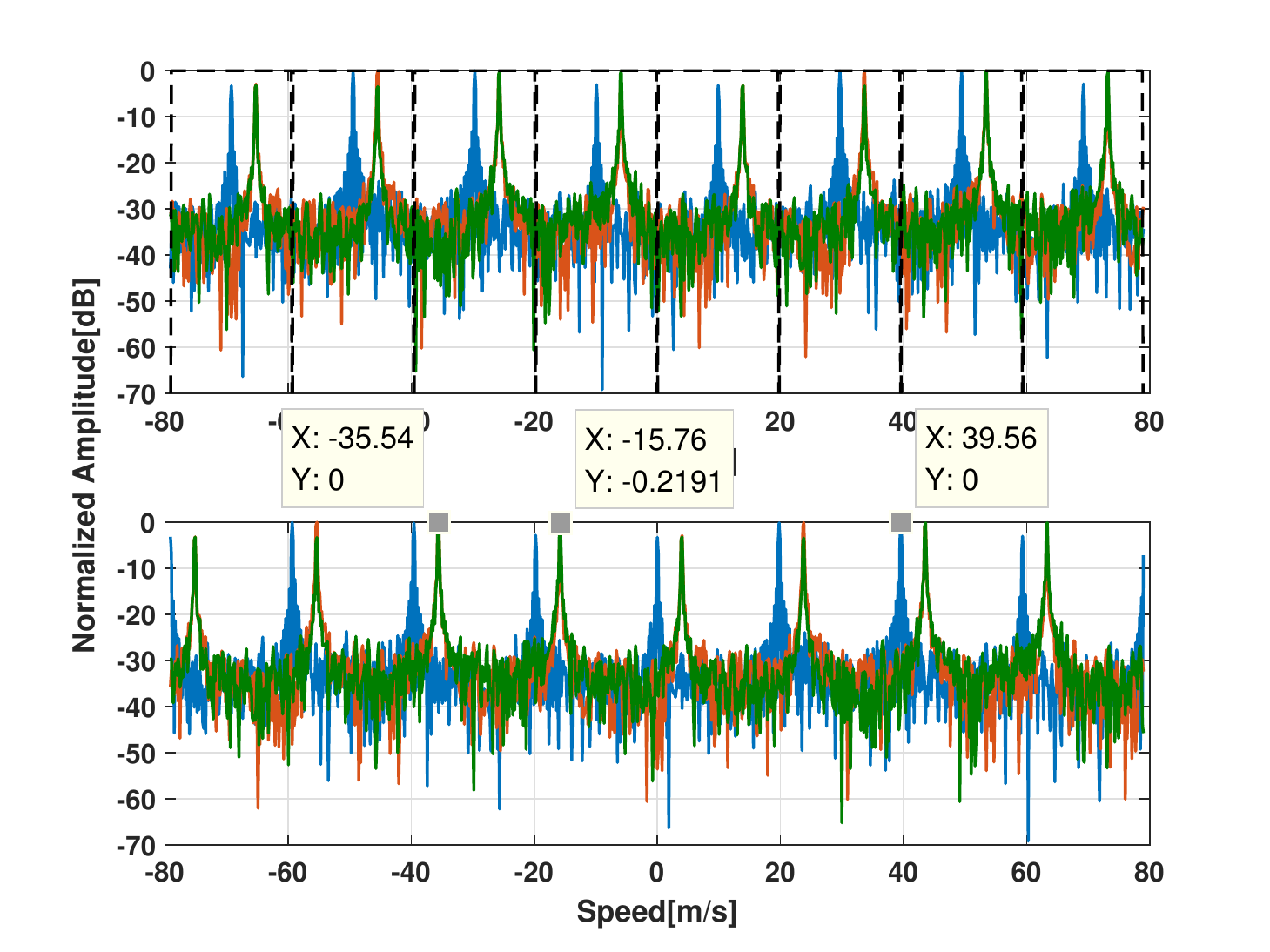}}
\caption{Signal processing results for TB DDMA approach, three targets.}
\label{fig4}
\end{figure*}

The conventional DDMA approach is capable of transmitting different phase modulated waveforms simultaneously, however, the requirement of unambiguous velocity reduces the chirp time and shortens the maximum unambiguous range. Such approach can only be used in SRR applications. If we increase the chirp time by $M$ times to compensate the maximum unambiguous range, the maximum unambiguous velocity for each transmit-receive channel after matched-filtering will merely be $\pm 10~{\rm m/s}$, which is unacceptable for automotive radar applications.

In this example, the FMCW signal has a chirp time $t_p = 12us$. Three targets are placed at $400 ~{\rm m}, 800 ~{\rm m}$ and $1200 ~{\rm m}$  with normalized Doppler shifts $1/4, -9/40$ and $-1/10$ (corresponding to velocities $40 ~{\rm m/s}, -35~{\rm m/s}$ and $-15~{\rm m/s}$), respectively. The other parameters are the same as in the previous example. The maximum detectable range is the same as that for the conventional TDMA scheme. If the conventional DDMA approach is used, the unambiguous Doppler spectrum for each transmit-receive channel is only $\pm 1/16$ (corresponding to velocity $\pm 10~{\rm m/s}$), which means that all three targets are folded across the Doppler domain and cannot be recognized correctly.

To tackle the above problem, the DDMA approach with empty Doppler spectrum is used. The proposed method aims at maintaining the maximum unambiguous velocity (i.e., $\pm 80~{\rm m/s}$) using the DDMA approach. Instead of uniformly dividing the Doppler spectrum into $M = 8$ pieces, we now design $M_{\rm v} = 16$ different phase modulation sequences. Only the first $M = 8$ sequences are selected to form the phase modulation matrix ${\bf W}$. The range-Doppler map of the received signal in a single receive element is shown in Fig.~\ref{fig3a}. As compared to Fig.~\ref{fig2a}, ${R_{max}}$ and $v_{max}$ are identical. Unlike their counterparts in the conventional DDMA approach, the peaks of different targets appear more dense across the Doppler index. Half of the Doppler spectrum is reserved. The empty Doppler spectrum behaves like an anchor that connects the repeated peaks in range-Doppler map for different pre-designed Doppler shifts.

To recover the unambiguous target velocity via the empty Doppler spectrum, the first step is to determine the potential locations of the targets from Fig.~\ref{fig3a}. A binary matrix that has the same size as the range-Doppler map is generated using \eqref{eq19} and \eqref{eq20}. Note that the range-Doppler map of each receive element is transformed into a sub binary matrix, the intersection of these sub binary matrices helps to avoid potential false alarms and improve the robustness of detection. As shown in Fig.~\ref{fig3b}, the binary matrix contains only 1's representing targets and 0's denoting noises. Particularly, 1's usually appear at the same range cell with different Doppler shifts, and the distances between those Doppler shifts are identical. Given that the default Doppler shifts are increasing, the first peak almost surely belongs to the first transmit element. The actual Doppler shift of the target can therefore be resolved. Since the Doppler spectrum is cyclic after FFT, we use the thresholding function \eqref{eq22} to recover the target Doppler shifts. The Doppler slices at target range cells after a proper demodulation are shown in Fig.~\ref{fig3c}, where the first peak of each target gives the correct Doppler shift.

By introducing the empty Doppler spectrum, the Doppler ambiguity problem in the conventional DDMA approach is solved. The maximum unambiguous velocity is fully recovered. As a result, the decline of the chirp time and the maximum detectable range is avoided. The maximum unambiguous range and velocity that the conventional TDMA scheme possess is then also achieved using the DDMA approach, while the loss of transmitted energy is also mitigated.

\subsection{Example~3: TB DDMA Approach}
Considering a typical working scenario of automotive radar, it is useful to focus the transmitted energy on a fixed spatial region via TB design. In this example, the TB DDMA approach is used to form a DFT-based TB in slow-time and achieve waveform diversity in Doppler domain. The parameters are the same as those in last example.

First, a virtual phase modulation matrix of size $M_{\rm v} \times Q_{\rm v}$ is generated using \eqref{eq16}. With $M_{\rm v} = 16$ optional phase modulation sequences, the DDMA waveform can generate 16 different beam directions thanks to virtual transmit beamforming. Only 8 of these 16 directions should be selected since the number of real transmit elements is $M = 8$. Assuming that the spatial region of interest is $\Theta : \; [-30^\circ, 30^\circ]$, all $M = 8$ beamforming vectors can be devised into two parts. The first 4 beamforming vectors cover the region $[0^\circ,30^\circ]$ while the other 4 vectors focus on the region $[-30^\circ,0^\circ]$. Therefore, the indices of selected pulses in every $M_{\rm v} = 16$ pulses are $1,2,3,4$ ($\sin\theta_{\rm v} = \{0,1/16, 1/8, 3/16\}$) and $13,14,15,16$ ($\sin\theta_{\rm v} = \{-1/4,-3/16,-1/8,-1/16\}$), respectively. If the same $M$ rows are chosen, the phase modulation matrix of the TB DDMA approach is regarded as a submatrix of the phase modulation matrix with empty Doppler spectrum after column selection. The received signal, in this case, is obtained by a pulse resampling from the received signal of the DDMA approach with empty Doppler spectrum, which essentially yields a decimation operator.

Consequently, the range-Doppler map of the received signal of a single receive element in the TB DDMA approach is shown in Fig.~\ref{fig4a}, where the entire Doppler spectrum is occupied by $M$ transmit elements. Note that the Doppler ambiguity mitigation can be conducted before the pulse resampling. The maximum unambiguous velocity in the TB DDMA approach is also $\pm 80~{\rm m/s}$. The chirp time of the FMCW signal can be $12us$ instead of only $1.5us$ to achieve the desired ${R_{max}}$, which is different from the conventional DDMA approach in Fig.~\ref{fig2c}. The binary matrix is shown in Fig.~\ref{fig4b}, which also proves that there is no empty Doppler spectrum in the TB DDMA approach. Using the recovered target Doppler shifts before pulse resampling, the slices of three targets are demodulated and shown in Fig.~\ref{fig4c}. It can be seen that the peaks of the last two targets are completely overlapped. The normalized Doppler shift difference between them is $1/8$ , which is also the step size of the pre-designed Doppler shift vector. The cyclic phase modulation mixes the Doppler slices of the last two targets, which also happens in the conventional DDMA approach. Two targets with particular Doppler shift difference may be mistaken for one single target, which is fatal for automotive radar applications. Again, this issue can be solved by the Doppler ambiguity mitigation method. The test function \eqref{eq21} may return some false alarms when targets~2 and~3 are in the same range cell. At most nine targets can be detected and only the first and the last one are true targets. However, the false alarms happen rarely and they do not degrade the radar detection ability in automotive radar applications.

\subsection{Example~4: Structure of the TB}
Two TB design methods in slow-time and fast-time are performed in this example. The same parameters as in the last simulation example are used here. The design of the TB matrix ${\bf D}$ in fast-time is conducted by solving optimization problem \eqref{eq14}, while the DFT-based TB is synthesized via the TB DDMA approach. For the fast-time TB design, we consider $K = 4$ waveforms to focus the energy within the region $\Theta : \; [-30^\circ, 30^\circ]$, while for the slow-time TB design, there are $M_{\rm v} = 24$ virtual transmit elements. Consequently, the beamforming vectors that point at $\sin\theta_{\rm v} = \{0,\pm 1/24, \pm 1/12, \pm,1/8, 1/6\}$ are selected. The phase modulation matrix is generated by selecting the corresponding pulses in every $M_{\rm v} = 24$ pulses.

\begin{figure}
\centerline{\includegraphics[width=\columnwidth]{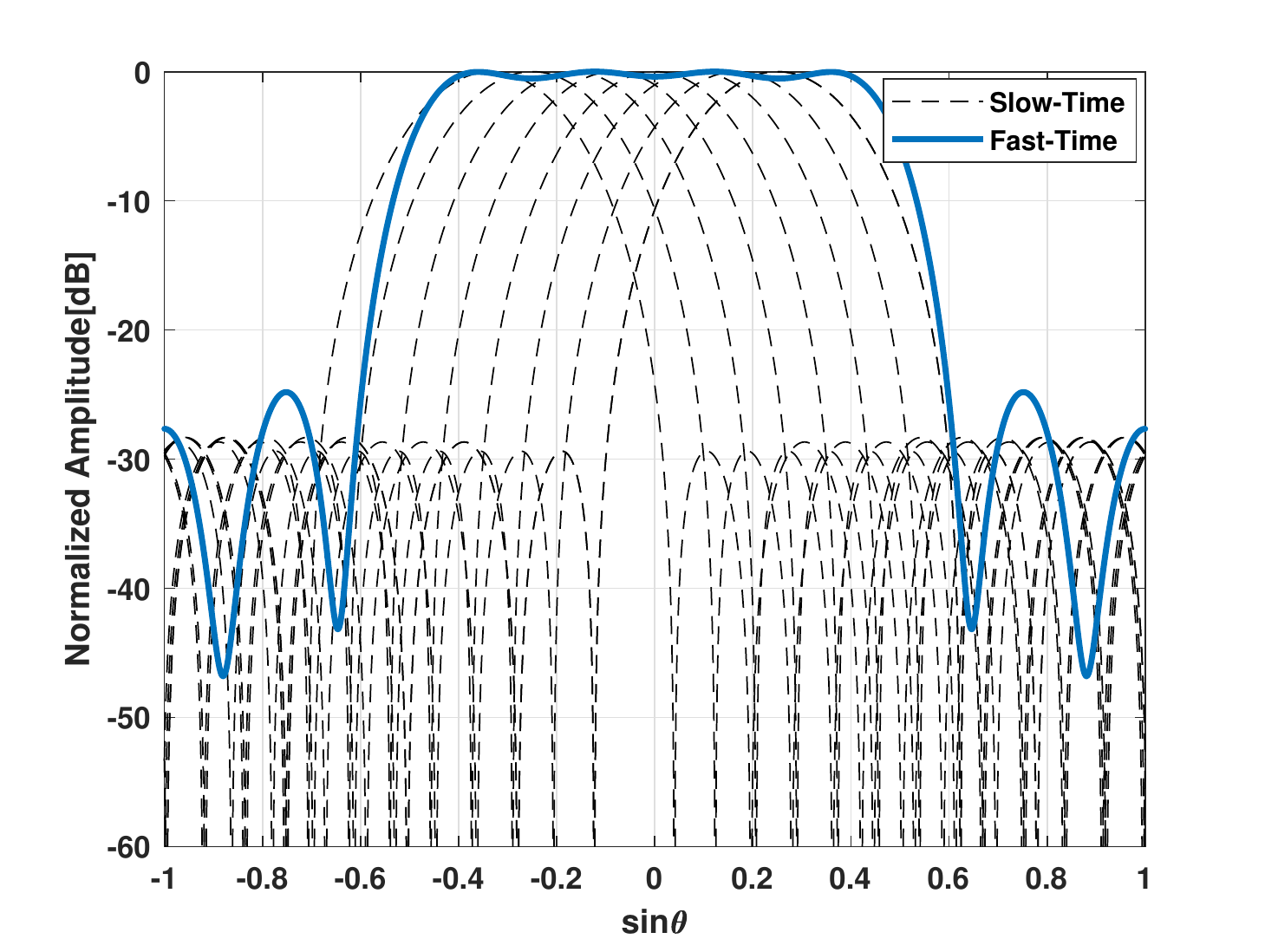}}
\caption{TB in fast-time and slow-time.}
\label{fig5}
\end{figure}

Fig.~\ref{fig5} illustrates the normalized beampatterns of the aforementioned TB design methods. Note that a Taylor window is applied on the beamforming vectors that form the DFT-based TB in slow-time, which has no influence on the waveform diversity in Doppler domain. Both methods yield an effective TB that can focus the transmitted energy for automotive MIMO radar.

\begin{figure}
\centerline{\includegraphics[width=\columnwidth]{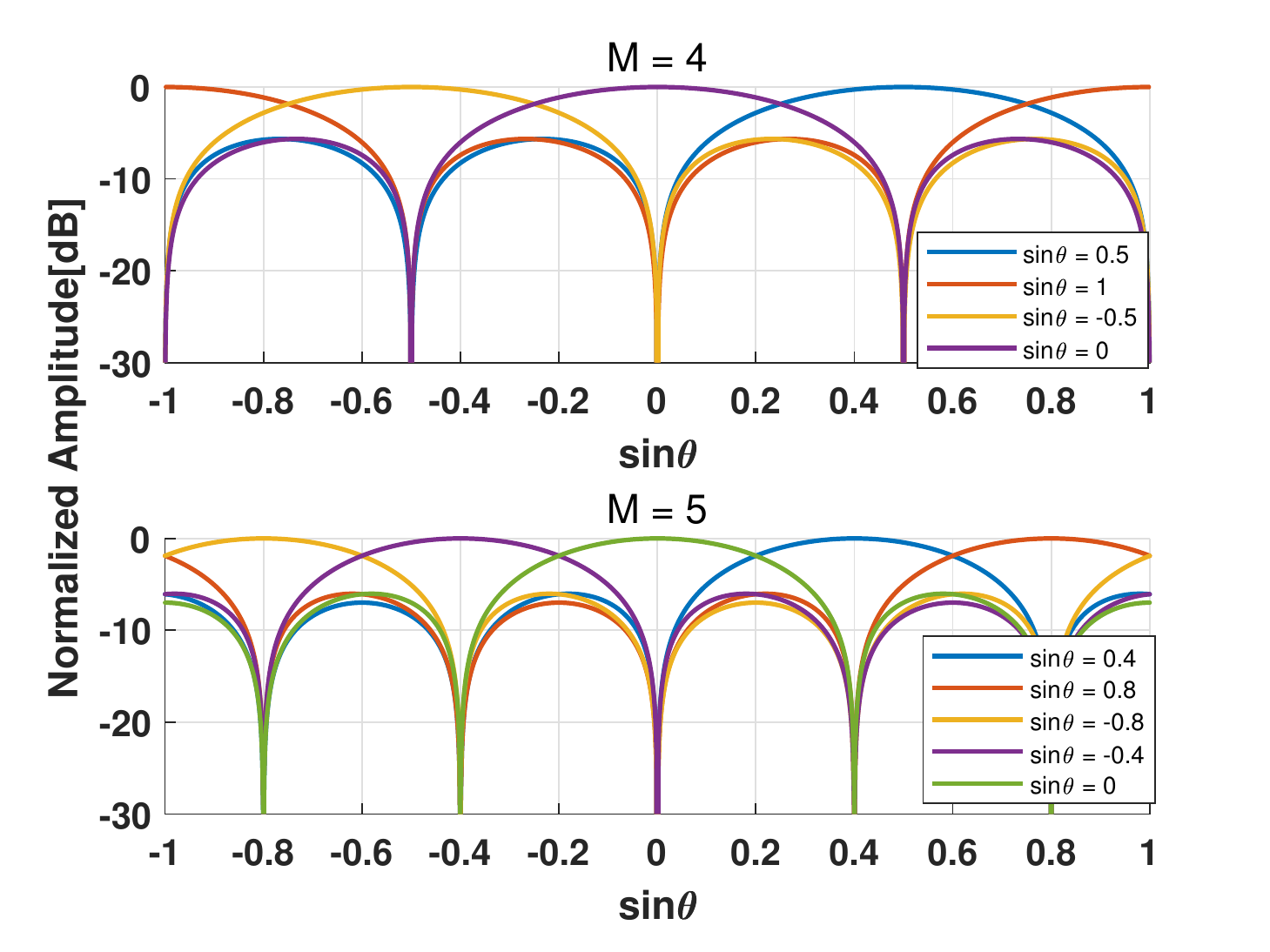}}
\caption{Virtual transmit beamforming of the DDMA approach.}\label{BeamScan}
\end{figure}

\section{Conclusion}
\begin{figure*}
    \centering
  \subfloat[Conventional DDMA.\label{PhaseVersusPulsea}]{%
       \includegraphics[width=0.33\linewidth]{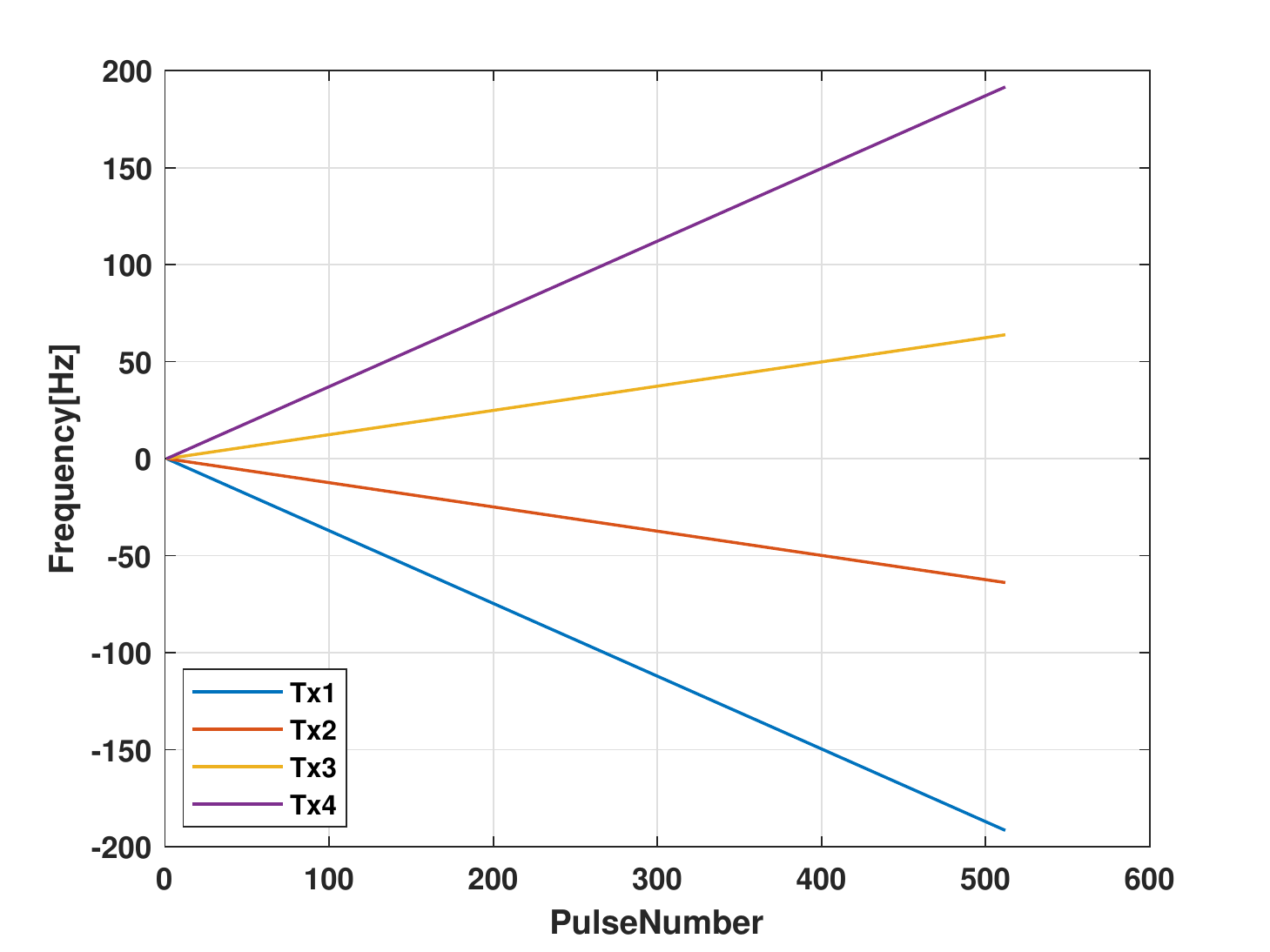}}
    \hfill
  \subfloat[DDMA with empty Doppler spectrum.\label{PhaseVersusPulseb}]{%
        \includegraphics[width=0.33\linewidth]{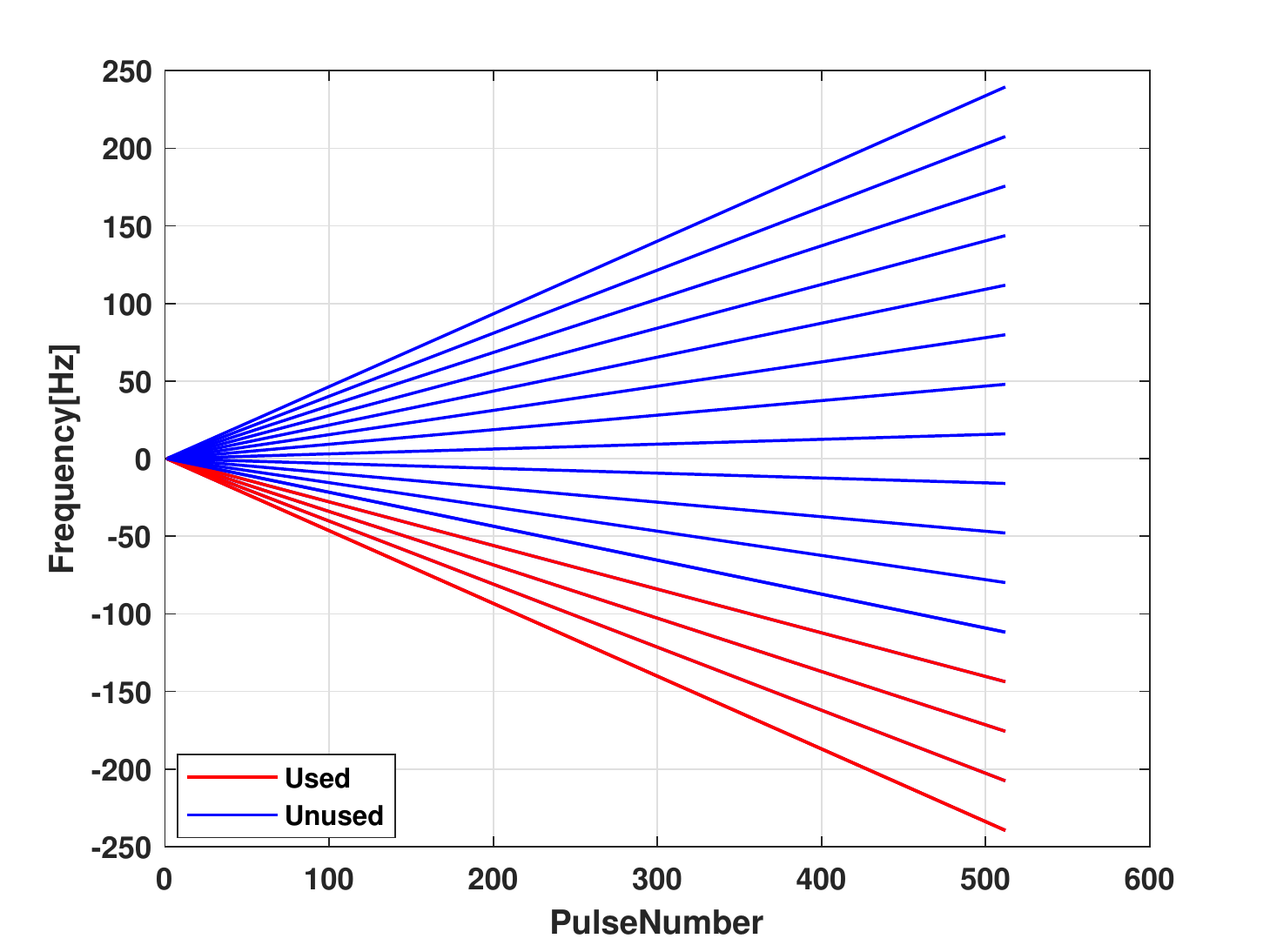}}
    \hfill
  \subfloat[TB DDMA.\label{PhaseVersusPulsec}]{%
        \includegraphics[width=0.33\linewidth]{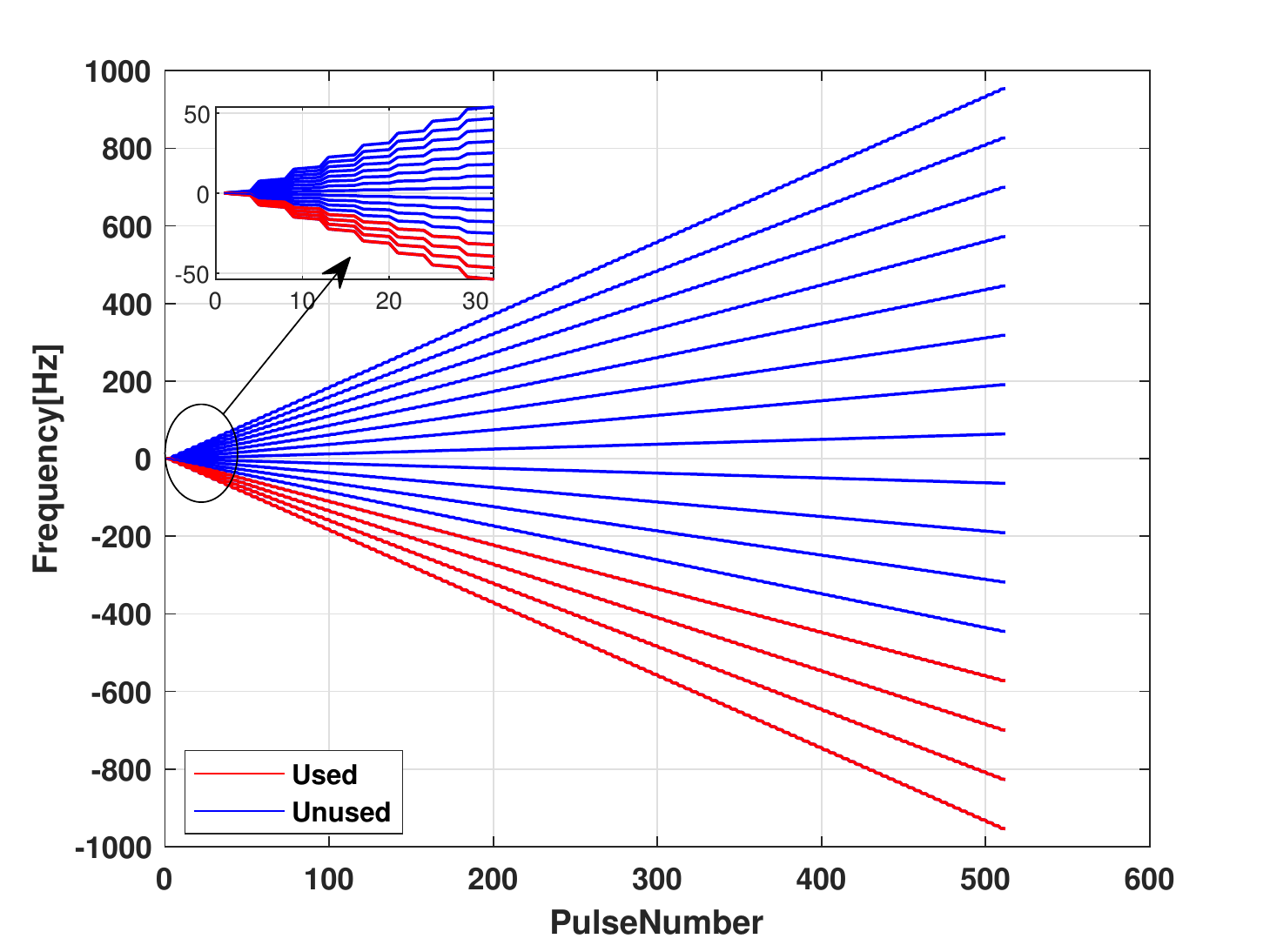}}
\caption{Different phase modulation matrix design approaches for DDMA waveform.}
\label{PhaseVersusPulse}
\end{figure*}

In this paper, the TB DDMA approach has been proposed to address the waveform design problem for automotive MIMO radar with improved transmit energy distribution. Particularly, the phase modulation matrix has been designed to enable a part of the Doppler spectrum as empty spectrum. A sequential detection method has been proposed to tackle the Doppler ambiguity problem via the use of the empty Doppler spectrum. By exploiting the virtual transmit beamforming associated with the DDMA waveform, the DFT-based TB has been generated to concentrate the transmitted energy within a fixed spatial region, which is a typical need in automotive radar applications. The proposed TB DDMA approach can form the TB in slow-time via pulse resampling at the receiver side while keeping the waveform diversity in Doppler domain. The proposed method requires no change of the transmitter structure and involves only basic algebra operators. As compared to the currently popular TDMA technique, our TB DDMA approach fully exploits the transmission capabilities of all transmit elements and avoids the trade-off between the active time for each transmit element and the frame time. Also, the proposed approach can be used in automotive MIMO radar with many transmit antennas since the DDMA code families can be generated easily. Simulation results have verified the effectiveness of the proposed TB DDMA based automotive MIMO radar method.

\appendices
\vspace{-3mm}
\section{Proof of \eqref{eq18}}\label{appendixA}

Insert the pre-designed $f_m$ into \eqref{eq16} and rewrite it as
\begin{equation}\label{ap1}
  {\bf w}_q = \left[1,e^{-j\pi\frac{2q}{M}},\cdots,e^{-j\pi\frac{2(M-1)q}{M}}\right]^T.
\end{equation}

If $f_m = 0.5 {f_a} \left( -1 + {(2m-1)}/{M} \right)$ is used, then \eqref{ap1} can be updated as
 \begin{equation}\label{ap2}
  {\bf w}_q = \xi_q \left[1,e^{-j\pi\frac{2q}{M}},\cdots,e^{-j\pi\frac{2(M-1)q}{M}}\right]^T
\end{equation}
where $\xi_q \triangleq e^{j\pi q(1-1/M)}$ is fixed during the pulse and has no influence on the beamformer direction. Note that  \eqref{ap2} is just the scaled version of \eqref{ap1}. Thus, there is no critical differences between these two coding methods.

It can be seen that \eqref{ap1} is a DFT weight vector with factor $2q/M$. Comparing it with the transmit steering vector of a ULA with $M$ elements and interelement spacing $d_t$, i.e., ${\bf a}(\theta_{\rm v}) \triangleq [1,e^{-j \frac{2\pi }{\lambda}d_t\sin\theta_{\rm v}},\cdots, e^{-j\frac{2\pi }{\lambda}d_t(M-1)\sin\theta_{\rm v}}]^T$, it can be found that $|{\bf a}(\theta_{\rm v})| = |{\bf w}_q|$ if and only if
\begin{equation}\label{ap3}
  \sin\theta_{\rm v} + 2j = \frac{q}{\mu M},\quad q = 1,2,\cdots, Q
\end{equation}
where $j \in \mathbb{Z}$ is a nonnegative integer, $\mu = d_t/\lambda$, and we usually let $\mu = 0.5$. Like in the TDMA scheme, we assume $Q = {\bar Q}M$. Then, it can be verified that the distinct virtual beamformer directions $\theta_{\rm v}$ are determined by $\sin\theta_{\rm v} = {2q^\star}/{M}$, where $q^{\star} \in \left( { - {{\rm{M}}}/{2}, {{\rm{M}}}/{2}} \right]$ contains $M$ different integers. These directions are symmetrically distributed on the entire spatial region $\sin\theta \in [-1,1]$. Note that the DDMA phase coding is recycled every $M$ pulses, and the beamformer direction also has a period of $M$ pulses accordingly. Therefore, all these directions are illumined $\bar Q$ times.

If we divide the phase modulation matrix for the DDMA approach into $\bar Q$ submatrices, i.e., ${\bf W} = {\left[ {{\bf \Omega}^{(1)},{\bf \Omega}^{(2)},\cdots,{\bf \Omega}}^{(\bar Q)} \right]}$, it can be observed that the submatrices ${\bf \Omega}^{(\bar q)} \in {\mathbb{C}^{M \times M}}, {\bar q} = 1,2,\cdots, {\bar Q}$ are identical, which implies that ${\bf W}$ for the DDMA approach has the same structure as in \eqref{eq6}. Consequently, the phase modulation matrix for the DDMA approach can be modelled as
\begin{equation}\label{ap5}
  {\bf W} = \underbrace {\left[ {{\bf \Omega}, {\bf \Omega},\cdots,{\bf \Omega}} \right]}_{\bar Q}
\end{equation}
where ${\bf \Omega} \triangleq [{\bf w}_1,{\bf w}_2,\cdots, {\bf w}_M]$ is a scaled orthogonal matrix containing $M$ beamforming vectors pointing at $\theta_{\rm v}$, and we drop the superscript for simplicity.

For example, let us consider $M = 4$. The directions of the first $2M$ pulses are given by
\begin{equation}\label{ap6}
  \sin {\theta _v} = \underbrace {0.5,1}_{j = 0},\quad \underbrace { - 0.5,0}_{j =  1}, \quad \underbrace {0.5,1}_{j = 2},\quad \underbrace { - 0.5,0}_{j =  3}
\end{equation}
which shows that the repeated cycle is $M = 4$. The same result can be observed if $M$ is odd (e.g., $M = 5$). Fig.~\ref{BeamScan} illustrates the beampattern generated by ${\bf \Omega}$ in one cycle with $M = 4$ and $M = 5$, respectively.

It is also worth noting that the virtual transmit beamforming vectors are orthogonal to each other, i.e., the peak of any particular beampattern is the null of the other beampatterns. This property is ensured by the vectors in ${\bf \Omega}$, where the inner product of any two beamforming vectors ${\bf w}_q$ is zero. The transmitted signals, although share the same envelope, are assigned to different directions with a unique orthogonal beampattern. The waveform diversity is achieved since there is no linear combination of transmitted signals in space. The overall transmit beampattern is still omnidirectional.

\section{Phase Values for Different DDMA Approaches}\label{appendixB}

For the conventional DDMA approach, the phases for each transmit antenna grow linearly at different rates. The rates are also uniform. The DDMA approach with empty Doppler spectrum uses only $M$ continuous sequences among the total of $M_{\rm v}$ code sequences, while the TB DDMA approach extracts particular pulses based on the DDMA approach with empty Doppler spectrum. Fig.~\ref{PhaseVersusPulse} gives an example with $M = 4$ and $M_{\rm v} = 16$ for $Q = 512$ pulses. Note that the phases vary nonlinearly in Fig.~\ref{PhaseVersusPulsec}, and there is a cyclic step due to pulse selection.

\balance
\bibliographystyle{IEEEtran}
\bibliography{IEEEabrv,Ref}

\end{document}